\documentclass[lettersize,journal]{IEEEtran}
\usepackage{amsmath,amsfonts}
\usepackage{algorithmic}
\usepackage{algorithm}
\usepackage{array}
\usepackage[table]{xcolor}  

\usepackage{textcomp}
\usepackage{stfloats}
\usepackage{url}
\usepackage{verbatim}
\usepackage{graphicx}
\usepackage{cite}
\usepackage{amsthm,amsmath,amssymb}
\usepackage{mathrsfs}
\usepackage{enumitem}
\usepackage{subcaption} 
\usepackage{times}
\usepackage{svg}
\usepackage{multirow}
\usepackage{amsthm}            
\usepackage{booktabs}

\theoremstyle{plain}
\newtheorem{theorem}{Theorem}

\newtheorem{definition}{Definition}
\newtheorem{corollary}{Corollary}

\usepackage{pifont}
\newcommand{\cmark}{\ding{51}}           
\newcommand{\xmark}{\ding{55}}           
\newcommand{\dmark}{\ensuremath{\triangle}} 

\begin{document}

\title{Joint Trajectory, RIS, and Computation Offloading Optimization via Decentralized Model-Based PPO in Urban Multi-UAV Mobile Edge Computing}

\author{Liangshun Wu~\IEEEmembership{Member,~IEEE}, Jianbo Du~\IEEEmembership{Senior Member,~IEEE}, and Junsuo Qu*~\IEEEmembership{Member,~IEEE}\\

\thanks{This research is funded by the Xi’an Key Laboratory of Advanced Control and Intelligent Process (Grant No. 2019220714SYS022CG04), the Key R\&D Plan of Shaanxi Province (Grant No. 2021ZDLGY04-04), Collaborative Innovation Project of Xi'an Science and Technology Bureau (Grant No. 24KGDW0022), {the open research fund of National Mobile Communications Research Laboratory, Southeast University (Grant No. 2026D03),   the National Natural Science Foundation of China (Grant 62271391)}, Shanghai Key Laboratory of Trustworthy Computing (East China Normal University)  (Grant No. 24Z670103399), Key Laboratory of Embedded System and Service Computing (Tongji University), Ministry of Education (Grant No. ESSCKF2024-10), and Key Laboratory of Computational Neuroscience and Brain-Inspired Intelligence (Fudan University), Ministry of Education (Grant No. 25Z670102051).}

\thanks{* Corresponding author. Email:qujunsuo@xupt.edu.cn.}
\thanks{Liangshun Wu is with {School of Communications and Information Engineering, Xi’an University of Posts \& Telecommunications, Xi’an 710061, China}, and School of Information and Electronic Engineering, Shanghai Jiao Tong University, Shanghai 200240, China (Email: wuliangshun@sjtu.edu.cn).
{Jianbo Du is with National Mobile Communications Research Laboratory, Southeast University. Jianbo Du is also with Shaanxi Key Laboratory of Information Communication Network and Security, School of Communications and Information Engineering, Xi'an University of Posts and Telecommunications, Xi'an 710121, China. } (Email: dujianboo@163.com). Junsuo Qu is with Xi'an Key Laboratory of Advanced Control and Intelligent Process, School of Automation, Xi’an University of Posts \& Telecommunications, Xi’an 710061 (Email: qujunsuo@xupt.edu.cn).}

}


\maketitle

\begin{abstract}
Efficient computation offloading in multi-UAV edge networks becomes particularly challenging in dense urban areas, where line-of-sight (LoS) links are frequently blocked and user demand varies rapidly. Reconfigurable intelligent surfaces (RISs) can mitigate blockage by creating controllable reflected links, but realizing their potential requires tightly coupled decisions on UAV trajectories, offloading schedules, and RIS phase configurations. This joint optimization is hard to solve in practice because multiple UAVs must coordinate under limited information exchange, and purely model-free multi-agent reinforcement learning (MARL) often learns too slowly in highly dynamic environments. To address these challenges, we propose a decentralized model-based MARL framework. Each UAV optimizes mobility and offloading using observations from several hop neighbors, and submits an RIS phase proposal that is aggregated by a lightweight RIS controller. To boost sample efficiency and stability, agents learn local dynamics models and perform short horizon branched rollouts for proximal policy optimization (PPO) updates. Simulations show near centralized performance with improved throughput and energy efficiency at scale.
\end{abstract}


\section{Introduction}
\IEEEPARstart{E}{nabling} low-latency and energy-efficient computation offloading is a critical objective for next generation wireless networks. Mobile edge computing (MEC) migrates computationally intensive tasks from devices to nearby edge servers, yet dense urban deployments often suffer line of sight (LoS) blockages and unreliable links due to buildings and obstacles. Integrating unmanned aerial vehicles (UAVs) with reconfigurable intelligent surfaces (RIS) has been shown to effectively mitigate these challenges by exploiting UAV mobility and programmable reflections to create virtual LoS paths, extend base station (BS) coverage, and improve energy efficiency and offloading reliability {\cite{yang2025energy,qin2023joint,naaz2024empowering,khan2021large,yang2021communication,bansal2023ris,wang2025survey}}. With the shift toward mmWave {or} THz bands, where coverage shrinks and signals are highly susceptible to blockage, UAV-RIS architectures have emerged as a flexible, cost aware means to fill coverage holes and support latency sensitive MEC~{\cite{yang2025energy,sun2024multi,naaz2024empowering,wu2025towards,jiang2024delay,wang2024real,saif2024effectiveness,prabhashana2025machine}. 
Recent works focus on optimizing energy efficiency, latency, and quality of experience (QoE)  in UAV RIS assisted MEC systems~\cite{wu2025towards,khalil2023deep,jiang2024delay,wang2024real,saif2024effectiveness,prabhashana2025machine,michailidis2024optimization,he2025qoe,nguyen2025ground,abdalla2025secrecy,chen2025energy,wang2025joint,wang2025survey}.}

\begin{table*}[h]
\centering
\caption{Representative MARL methods under partial observability and sample-efficiency constraints.}
\renewcommand{\arraystretch}{1.25}
\arrayrulecolor{black}
\begin{tabular}{p{3.3cm}|p{2.8cm}|p{2.8cm}|p{2.8cm}|p{3.4cm}}
\hline
{Feature / Method} &
{Comm.-Efficient MARL~\cite{chu2020multi,wang2025lns2+,qu2020scalable}} &
{Neighbor Comm. ~\cite{singh2018learning}} &
{Model-Based MARL~\cite{han2019grid,deisenroth2011pilco,wu2023models}} &
{Proposed MB-DRL (Ours)} \\ 
\hline
{Partial observability handling} & {\cmark (local info only)} & {\cmark (neighbor aggregation but fail to capture complex, coupled dynamics)} & {\dmark (centralized critic)} & {\cmark\cmark (decentralized $\kappa$-hop)} \\ 
\hline
{Communication cost} & {\cmark (compressed / truncated updates)} & {\cmark (limited to neighbors)} & {\xmark (global exchange)} & {\cmark\cmark (local $\kappa$-hop only)} \\
\hline
{Dynamics modeling} & { \xmark (model-free)} &{\xmark (model-free) }& {\cmark (learned transition) }& {\cmark\cmark (localized predictive model)} \\
\hline
{Sample efficiency} & {\xmark} &{\xmark} & {\cmark} & {\cmark\cmark (branched rollouts)} \\
\hline
{Convergence guarantee} & {\xmark} & {\xmark} & {\dmark (weak theoretical bounds)} & {\cmark (bounded model error)} \\
\hline
{Scalability to large UAV nets} & {\dmark (requires sync)} & {\cmark (distributed)} & {\xmark (centralized critic)} & {\cmark\cmark (fully decentralized)} \\
\hline
\end{tabular}
\vspace{1ex}
\begin{minipage}{0.95\textwidth}
\footnotesize
{\textit{Note:} 
\cmark\cmark = significant advantage; 
\cmark = available; 
\dmark = limited; 
\xmark = not supported.}
\end{minipage}
\arrayrulecolor{black}
\label{tab:com}
\end{table*}

{Traditional convex optimization and {model predictive control} (MPC) methods rely on accurate mathematical models \cite{morari1988model} and convex assumptions, while heuristic algorithms (e.g. genetic algorithm) depend on handcrafted rules, making them unsuitable for highly dynamic and partially observable RIS–UAV–MEC environments. In contrast, multi-agent reinforcement learning (MARL) learns adaptive and scalable control policies directly from interactions, effectively handling nonlinear coupled dynamics, partial observability, and decentralized decision-making\cite{wu2025towards,khalil2023deep,jiang2024delay,wang2024real}. However, since MARL algorithms rely on neural network approximations, the training process involves non-convex optimization, where convergence can be difficult to achieve or even diverge~\cite{wang2020data}. As system scale and complexity grow,  it poses challenges on existing MARL methods:
\begin{enumerate}
\item Partial observability. Most MARL methods (e.g. MADDPG)  follow centralized training and decentralized execution (CTDE), relying on a global critic to stabilize training but requiring costly information exchange; fully decentralized alternatives avoid this overhead yet often suffer from higher variance and weaker coordination~\cite{liu2024cooperative}. 
\item Low sample efficiency. The reliance of model-free MARL on costly interactions impedes its practical deployment in large-scale RIS-UAV-MEC systems~\cite{wang2025survey,hairi2024sample}.
\end{enumerate}}

Existing studies address these challenges through approaches that reduce communication costs and incorporate learned models. Under communication constraints, agents often share information only with neighboring peers. For instance, I3CNet~\cite{singh2018learning} uses an aggregation function to combine local messages. In addition, truncated policy gradient and Q learning variants can update policies locally, thereby lowering communication overhead~\cite{singh2018learning,wang2025lns2+,qu2020scalable,simao2019safe}. Nevertheless, many existing methods rely on simplified linear assumptions~\cite{zhang2018fully} or i.i.d. assumptions~\cite{du2022scalable}, which limits their ability to represent nonlinear dynamics. Sample efficiency can be improved by model based MARL, which increases data utilization by learning environment dynamics or opponent models~\cite{han2019grid,deisenroth2011pilco,wu2023models}, {with particular relevance to aerial and terrestrial UAV networks~\cite{10857476,10195219,10382630}. Diffusion based generative solvers have also been explored for MEC optimization~\cite{11018297,11006143},} yet they typically require centralized critics, lack rigorous theoretical guarantees on model error bounds and convergence, and do not fully couple model learning with policy optimization~\cite{guo2024decentralized}. Other techniques, such as prioritized replay, remain largely restricted to tabular settings and are difficult to extend to continuous control tasks.

We propose a model-based decentralized RL framework (MB-DRL). This setting involves high dimensional continuous control and noisy learning signals under partial observability, where unconstrained policy updates can be unstable. We therefore employ proximal policy optimization (PPO) to limit policy changes via a clipped surrogate objective and support parallel per-agent training. Our contributions are:
\begin{enumerate}
\item To handle partial observability and communication constraints, each UAV acts as an autonomous agent that makes decisions using only local observations and limited $\kappa$-hop neighbor information, enabling scalable cooperation without centralized coordination. To overcome the linear approximation limitation~\cite{zhang2018fully}, our method employs a nonlinear representation that jointly encodes local states, neighbor policies, and hidden features through deep LSTM-based fusion. Unlike baselines (e.g., I3CNet\cite{singh2018learning}) that process only local or mean-field information, our design integrates multi-source contextual cues via concatenated nonlinear transformations of neighborhood states. This enables each agent to capture complex, coupled dynamics across UAVs, RIS, and MEC nodes.
\item To address the reliance on centralized critics and the lack of theoretical guarantees~\cite{guo2024decentralized}, our framework adopts a fully decentralized PPO structure in which each agent performs local policy and value updates using  neighbors' states. We integrate model learning directly with policy optimization, where locally learned dynamics guide short-horizon rollouts that refine policy gradients and stabilize learning. Furthermore, we provide a bounded error convergence analysis that explicitly accounts for both model uncertainty and policy uncertainty. Unlike tabular prioritized replay, our method adopts a continuous, experience based sampling scheme that is compatible with neural function approximation, thereby improving sample reuse and training stability.
\end{enumerate}

As summarized in Table~\ref{tab:com}, our proposed MB-DRL framework achieves full decentralization and higher sample efficiency compared with existing methods.

The remainder of this paper is organized as follows: Section II presents the system model and problem formulation of the UAV- and RIS-assisted MEC system. Section III details the proposed decentralized, model-based MARL algorithm, including local dynamics learning, branched rollouts, and decentralized PPO optimization. Section IV discusses experimental settings and performance evaluation. Finally, Section V concludes the paper and outlines future research directions.

\renewcommand{\arraystretch}{1.2}
\begin{table}[h!]
\centering
\caption{{List of Main Symbols}}
\begin{tabular}{p{2.5cm}p{5.5cm}}
\hline
{Symbol} & {Description} \\
\hline
{$\mathcal{U}, \mathcal{K}$} & {Sets of UAVs and user equipments (UEs)} \\
{$u, k, n$} & {UAV, UE, and time-slot indices} \\
{$N, \delta_t, T$} & {Number of time slots, slot duration, and total period} \\
{$\mathbf{q}_u[n] = [x_{u,n}, y_{u,n}, z_{u,n}]^T$} & {3D position of UAV $u$ at slot $n$} \\
{$\mathbf{V}_u[n],\, \mathbf{A}_u[n]$} & {Velocity and acceleration vectors of UAV $u$} \\
{$V_{\max}$} & {Maximum UAV speed} \\
{$\mathbf{q}_{u,0},\, \mathbf{q}_{u,F}$} & {Initial and final positions of UAV $u$} \\
{$\mathbf{\Theta}[n]$} & {RIS phase-shift matrix at slot $n$} \\
{$\theta_m[n]$} & {Reflection phase of the $m$-th RIS element} \\
{$M = M_y \times M_z$} & {Total number of RIS elements (URA configuration)} \\
{$\mathbf{h}_{ur}[n], \mathbf{h}_{ra}[n], \mathbf{h}_{ua}[n]$} & {Channels of UAV–RIS, RIS–AP, and UAV–AP links} \\
{$\bar{\mathbf{h}}_{ua}[n]$} & {Effective UAV–RIS–AP channel} \\
{$\beta_0$} & {Reference path loss at 1 m} \\
{$\alpha_i, K_i$} & {Path-loss exponent and Rician $K$-factor of link $i$} \\
{$d_i[n]$} & {Link distance at slot $n$} \\
{$B$} & {System bandwidth} \\
{$p_t,\, p_j$} & {Transmit powers of UE and jammer} \\
{$\mathbf{w}_{u,k}[n]$} & {Transmit beamforming vector for UAV $u$, UE $k$} \\
{$\mathbf{u}_{u,k}[n]$} & {Receive beamforming vector for UAV $u$, UE $k$} \\
{$\gamma_{u,k}[n],\, \gamma_{a,u,k}[n]$} & {SNR of UE–UAV and UAV–AP links} \\
{$\tau_{u,k}^o[n],\, \tau_{u,k}^R[n]$} & {Time allocations for offloading and relay phases} \\
{$l_{u,k}^{o}[n]$} & {Offloaded bits from UE $k$ to UAV $u$} \\
{$l_{a,u,k}^{R}[n]$} & {Relayed bits from UAV $u$ to AP} \\
{$l_{u,k}^{\text{loc}}[n],\, l_{u,k}^{\text{comp}}[n]$} & {Bits computed locally (UE) and by UAV $u$} \\
{$L_k$} & {Total task size of UE $k$ (bits to be processed)} \\
{$c_u, c_a$} & {CPU cycles required per bit at UAV and AP} \\
{$F_u^{\max}, F_a^{\max}$} & {Maximum CPU frequency of UAV and AP} \\
{$E_k^{\text{tx}}[n],\, E_k^{\text{comp}}[n]$} & {Transmission and computation energy of UE $k$} \\
{$E_u^{\text{comp}}[n],\, E_u^{\text{fly}}[n]$} & {Computation and propulsion energy of UAV $u$} \\
{$\vartheta_1, \vartheta_2$} & {Propulsion power coefficients (drag, hover loss)} \\
{$\mu_k,\, \mu_u$} & {Effective switched-capacitance coefficients (DVFS model)} \\
{$E[n]$} & {Total system energy consumption at slot $n$} \\
{$\pi^{\theta_u},\, V^{\phi_u}$} & {Policy and value networks of UAV $u$} \\
{$\hat{p}_u$} & {Learned local transition (dynamics) model} \\
{$\mathcal{D}^E_u,\, \mathcal{D}^M_u$} & {Environment and model replay buffers} \\
{$\mathcal{N}_u^{(\kappa)}$} & {$\kappa$-hop neighbor set of UAV $u$} \\
{$r_u[n]$} & {Reward of UAV $u$ at time slot $n$} \\
{$\beta$} & {Weight controlling throughput–interference or entropy regularization} \\
{$T$} & {Rollout horizon for model-based simulation} \\
{$\xi$} & {Upper bound of model error tolerance} \\
{$z$} & {Vector of all optimization variables} \\
\hline
\end{tabular}
\label{tab:symbols}
\end{table}

Main notations used in context are summarized in Table~\ref{tab:symbols}.

\section{System Model}

\begin{figure*}[h]
\centering
\includegraphics[width=\textwidth]{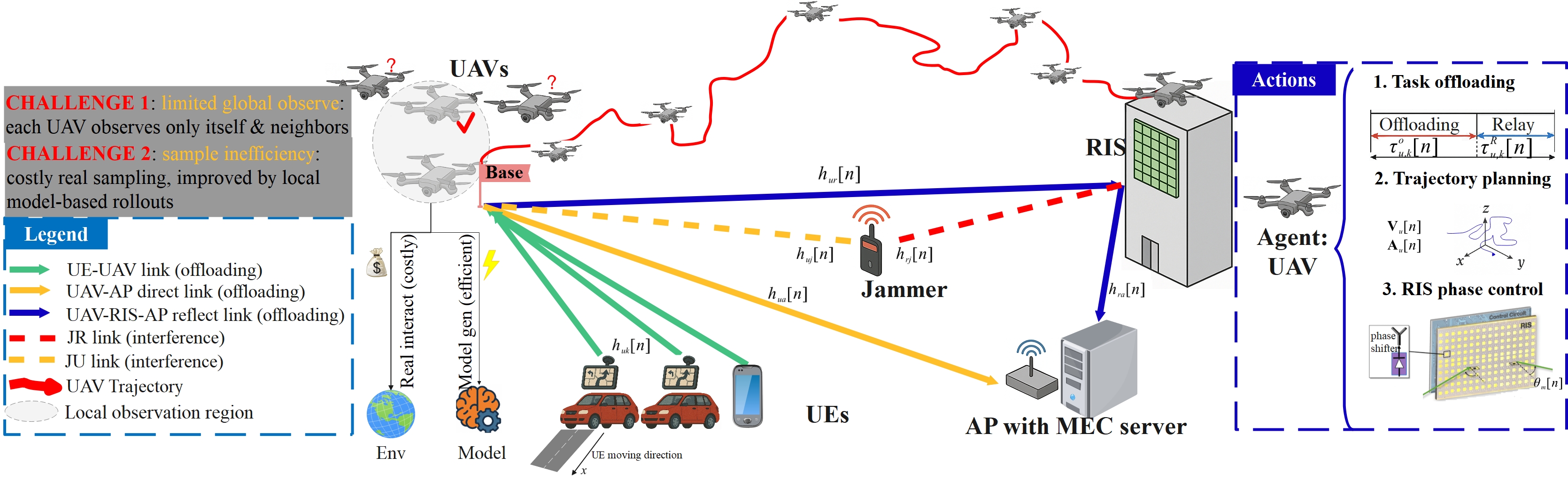} %
\caption{This work considers an RIS-assisted multi-UAV MEC system in which multi-antenna UAVs serve as both computing nodes and decode-and-forward relays for single-antenna UEs. UE tasks are offloaded over wireless links and either processed onboard or forwarded to a ground MEC server via direct and RIS reflected channels. Each UAV acts as an RL agent that jointly optimizes its trajectory, offloading decisions, and RIS phase recommendations, while a lightweight RIS controller aggregates these recommendations to enable coordinated beamforming. The  objective is to maximize system energy efficiency under two practical constraints: limited observability and low sample efficiency.}\label{scene}
\end{figure*}
\subsection{Scenario Setup}
To address the relay transmission problem in MEC  with UAVs and RIS, we consider a system where multiple UAVs (each with $L \geqslant 2$ antennas) serve both as computational nodes and relays, offloading tasks from multiple single-antenna user equipments (UEs) to a ground-based access point (AP) equipped with an MEC server. A building-mounted RIS, structured as a URA with $M_y \times M_z$ elements, reflects UAV signals toward the AP, as illustrated in Fig. \ref{scene}. Direct links from UEs to the RIS and AP are assumed to be blocked or highly attenuated due to obstructions and distance, making RIS-assisted reflection essential for improved service quality. The total transmission period $T$ is divided into $N$ uniform time slots $\delta_t$, with slot set $\mathcal{N}$ ($|\mathcal{N}| = N$).

In 3D space, the position of UE $k$ at time slot $n$ is denoted by $(w_k[n], 0)$, where $w_k[n] = (x_k[n], y_k)$, and $x_k[n] = x_k[0] + n$, representing the fact that each UE moves one unit along the $x$-direction per time slot.  The AP is located at $(w_a, 0)$ with $w_a = (x_a, y_a)$, the RIS at $(w_r, z_r)$ with $w_r = (x_r, y_r)$ and $z_r$ as its height, and an interferer/jammer at $(w_j, 0)$ with $w_j = (x_j, y_j)$. Only single-bounce RIS reflections are considered due to path loss, all channels are quasi-static flat fading, and full channel state information (CSI) is assumed available for system optimization.

Incorporating a jammer into the system model reflects practical scenarios where communication reliability must be maintained under intentional interference, e.g., illegal jammers near critical infrastructure in urban emergency networks, adversarial jamming in disaster relief or battlefield operations, and proactive interference defense in smart-city or cellular deployments. Explicitly modeling jamming allows us to evaluate robustness and provides a basis for designing anti-jamming strategies to ensure secure and reliable service.

\subsection{UAV Trajectory}

Each UAV operates over a limited duration of $N \delta_t$ seconds, which is divided into $N$ discrete time slots indexed by $n = 1, 2, \ldots, N$. Each slot spans $\delta_t$ seconds—sufficiently short to approximate the UAVs as static within a single slot. UAV $u \in \mathcal{U}$ travels from a known starting position $\mathbf{q}_{u,0} = [x_{u,0}, y_{u,0}, z_{u,0}]^T$ to a designated final destination $\mathbf{q}_{u,F} = [x_{u,F}, y_{u,F}, z_{u,F}]^T$, with its position at any time slot $n$ given by $\mathbf{q}_u[n] = [x_{u,n}, y_{u,n}, z_{u,n}]^T=(w_u[n],z_u[n])$ with $w_u[n]=(x_{u,n},y_{u,n})$.

We define the velocity and acceleration of UAV $u$ at time slot $n$ as $\mathbf{V}_u[n] = [V_{u,x}[n], V_{u,y}[n], V_{u,z}[n]]^T$ and $\mathbf{A}_u[n] = [A_{u,x}[n], A_{u,y}[n], A_{u,z}[n]]^T$, respectively. The UAV’s movement is governed by the following set of kinematic constraints:
\begin{equation}
\mathbf{q}_u[n+1]=\mathbf{q}_u[n]+\mathbf{V}_u[n] \delta_t+\frac{1}{2} \mathbf{A}_u[n] \delta_t^2, \quad \forall u \in \mathcal{U}, \forall n 
\label{eq1_multi}
\end{equation}
\begin{equation}
\mathbf{q}_u[0]=\mathbf{q}_{u,0}, \quad \mathbf{q}_u[N]=\mathbf{q}_{u,F}, \quad \forall u \in \mathcal{U}
\label{eq2_multi}
\end{equation}
\begin{equation}
\mathbf{V}_u[n+1]=\mathbf{V}_u[n]+\mathbf{A}_u[n] \delta_t, \quad \forall u \in \mathcal{U}, \forall n , n \leq N
\label{eq3_multi}
\end{equation}
\begin{equation}
\|\mathbf{q}_u[n]-\mathbf{q}_u[n-1]\| \leq \delta_t V_{\text{max}}, \quad \forall u \in \mathcal{U}, n=2,3, \ldots, N
\label{eq4_multi}
\end{equation}
Here, $V_{\text{max}}$ represents the UAV’s top allowable speed. According to the above, each UAV's future position is a function of its current state and motion dynamics. The initial and final positions are fixed at the first and last time slots, respectively. Velocity is incrementally updated based on acceleration, and each UAV’s displacement per slot must not exceed its maximum travel distance, ensuring feasible mobility.

\subsection{RIS Phase Control}
The RIS is mounted on a building wall parallel to the $xz$-plane and consists of $M = M_y \times M_z$ elements arranged in a URA. {The reflection phase coefficient matrix of the RIS at time slot $n$ is obtained by averaging the phase recommendations of all UAVs: 
\begin{equation}
\mathbf{\Theta}[n]=\bar{\mathbf\Theta}[n]=\frac{\sum_{u=1}^U \mathbf\Theta_u[n]}{U}.
\end{equation}
Specifically,
\begin{equation}
\boldsymbol{\Theta}_u[n]=\operatorname{diag}\left(e^{j \theta_{u,1}[n]}, e^{j \theta_{u,2}[n]}, \ldots, e^{j \theta_{u,M}[n]}\right),
\label{eq:snr_ap}
\end{equation}
represents the phase parameter vector recommended by UAV $u$ at time slot $n$ through UAV-RIS link, where each $\theta_{u,m}[n] \in [0, 2\pi)$ controls the phase of the $m$-th RIS element to enable intelligent beam steering.}\footnote{
{The aggregation minimizes communication overhead by updating synchronously only at selected time slots, reducing signaling while preserving performance.}}

\subsection{Channel Modeling}
\subsubsection{UAV-RIS-AP Link}

For UAV $u \in \mathcal{U}$ and the AP, the effective uplink channel incorporating both direct and RIS-assisted paths is modeled as:
\begin{equation}
\bar{\mathbf{h}}_{ua}[n] = \mathbf{h}_{ua}^{H}[n] + \mathbf{h}_{ra}^{H}[n] \mathbf{\Theta}[n] \mathbf{h}_{ur}[n],
\end{equation}
where $\mathbf{h}_{ua}^{H}[n] \in \mathbb{C}^{1 \times L}$ is the direct channel from UAV $u$ to the AP, $\mathbf{h}_{ur}[n] \in \mathbb{C}^{M \times L}$ is the LoS dominated channel from UAV $u$ to the RIS, $\mathbf{h}_{ra}^{H}[n] \in \mathbb{C}^{1 \times M}$ is the LoS channel from RIS to the AP.

$\mathbf{h}_{ur}[n]$ is modeled as:
\begin{equation}
\mathbf{h}_{ur}[n] = \frac{\beta_0}{\|\mathbf{q}_u[n] - \mathbf{p}_r\|} \mathbf{a}_{ru}^{R}[n] \mathbf{a}_u^{T}[n],
\end{equation}
where $\beta_0$ is the reference path loss at 1 meter, $\mathbf{p}_r = [x_r, y_r, z_r]^T$ is the position of the RIS, $\mathbf{a}_u^{T}[n] \in \mathbb{C}^{L \times 1}$ is the UAV's transmit array response vector,    
\begin{equation}
\mathbf{a}_u^{T}[n] = \left[1, e^{-j \frac{2\pi}{\lambda} d \cos\theta_{ur}[n]}, \ldots, e^{-j \frac{2\pi}{\lambda} (L-1) d \cos\theta_{ur}[n]} \right]^T,
\end{equation}
where $\theta_{ur}[n]$ is the angle of departure (AoD) from the UAV to the RIS in the elevation domain. $\mathbf{a}_{ru}^{R}[n] \in \mathbb{C}^{1 \times M}$ is the RIS receive array response vector.
\begin{equation}
\begin{aligned}
\mathbf{a}_{ru}^{R}[n] =& \mathbf{a}_{ry}^{R}[n] \otimes \mathbf{a}_{rz}^{R}[n] \in \mathbb{C}^{1 \times M}, \\
\mathbf{a}_{ry}^{R}[n] =& [1, e^{-j \frac{2\pi}{\lambda} d_y \sin \theta_{ur}[n] \cos \phi_{ur}[n]}, \\
                        &\ldots, e^{-j \frac{2\pi}{\lambda} (M_y - 1) d_y \sin \theta_{ur}[n] \cos \phi_{ur}[n]}], \\
\mathbf{a}_{rz}^{R}[n] =& [1, e^{-j \frac{2\pi}{\lambda} d_z \sin \theta_{ur}[n] \sin \phi_{ur}[n]}, \\
                        &\ldots, e^{-j \frac{2\pi}{\lambda} (M_z - 1) d_z \sin \theta_{ur}[n] \sin \phi_{ur}[n]} ].
\end{aligned}
\end{equation}
where $d_y$ is the spacing between adjacent RIS elements along the $y$-axis, and $(M_y-1)d_y$ represents the maximum physical span of the RIS array in the $y$-direction. So do $d_x$ and $(M_x-1)d_x$. The array response vector uses these spacings to model the phase shifts experienced by a plane wave arriving at different elements due to their spatial separation. 

The spatial angles are defined as:
\begin{equation}
\begin{aligned}
\sin \theta_{ur}[n] &= \frac{z_r - z_u[n]}{\|\mathbf{q}_u[n] - \mathbf{p}_r\|}, \\
\cos \phi_{ur}[n] &= \frac{y_r - y_u[n]}{\sqrt{(x_r - x_u[n])^2 + (y_r - y_u[n])^2}}, \\
\sin \phi_{ur}[n] &= \frac{x_r - x_u[n]}{\sqrt{(x_r - x_u[n])^2 + (y_r - y_u[n])^2}}.
\end{aligned}
\end{equation}

\begin{figure}[h]
\centering
\includegraphics[width=0.35\textwidth]{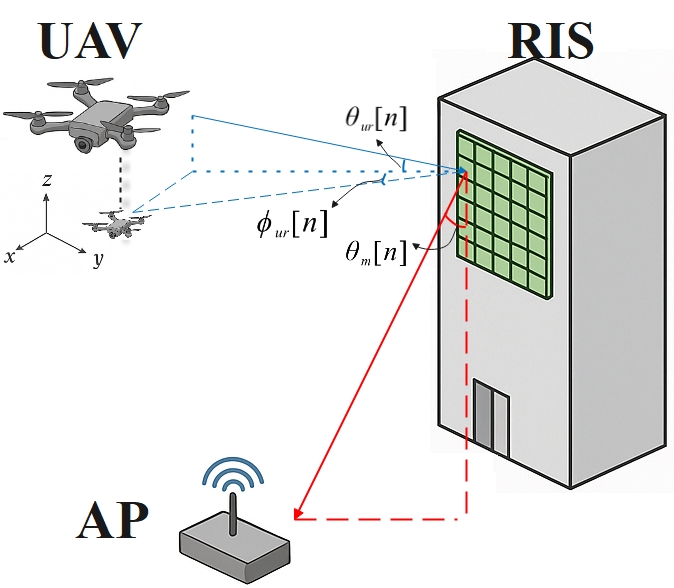} %
\caption{Geometric meaning of angles. The phase shift $\theta_m[n]$ applied by each RIS element is designed to align the reflected signal toward the AP, effectively corresponding to the angle of departure (AoD) from the RIS element to the AP. The angle $\theta_{ur}[n]$ denotes the angle of departure (AoD) from the UAV to the RIS element, representing the elevation direction of the incoming wave. The azimuth angle $\phi_{ur}[n]$ characterizes the angle of arrival (AoA) of the UAV’s signal at the RIS element projected onto the $xy$-plane. }\label{angle}
\end{figure}

Fig. \ref{angle} gives a schematic diagram of the geometric meaning of the UAV-RIS-AP link.

\subsubsection{UE-to-UAV, RIS-to-AP, and UAV-to-AP Links (Rician Fading)}

All other channels, including UE-to-UAV, RIS-to-AP, and UAV-to-AP, are modeled using a Rician fading model:
\begin{equation}
\begin{aligned}
\mathbf{h}_{i}[n] &= \sqrt{\beta_0 d_i^{-\alpha_i}[n]} \left( \sqrt{\frac{K_i}{1+K_i}} \mathbf{h}_{i}^{\text{LoS}}[n] + \sqrt{\frac{1}{1+K_i}} \mathbf{h}_{i}^{\text{NLoS}}[n] \right), \\
& i \in \{(u,k), (r,a), (u,a)\},
\end{aligned}
\end{equation}
where $\beta_0$ is the reference path loss at 1 meter, $K_i$ is the Rician $K$-factor for link $i$, $\alpha_i$ is the path-loss exponent for link $i$, $d_i[n]$ is the Euclidean distance of link $i$ at time slot $n$. $\mathbf{h}_{i}^{\text{LoS}}[n]$ is the deterministic line-of-sight component. For links involving antenna arrays (e.g., UAV or AP), this is modeled as an array response vector depending on the angle of arrival (AoA) or angle of departure (AoD):
\begin{equation}
    \mathbf{h}_{i}^{\text{LoS}}[n] = \mathbf{a}_{i}(\theta_i[n]),
\end{equation}
where $\mathbf{a}_{i}(\theta_i[n])$ is the corresponding array steering vector.
$\mathbf{h}_{i}^{\text{NLoS}}[n]$ is the non-line-of-sight component, modeled as:
\begin{equation}
    \mathbf{h}_{i}^{\text{NLoS}}[n] \sim \mathcal{CN}(0, \mathbb{I}),
\end{equation}
representing Rayleigh fading with i.i.d. complex Gaussian entries.

The corresponding link distances are:
\begin{equation}
\begin{aligned}
d_{uk}[n] &= \sqrt{ \|{w}_u[n] - {w}_k[n]\|^2 + z_u[n]^2 }, \\
d_{ua}[n] &= \sqrt{ \|{w}_u[n] - {w}_a\|^2 + z_u[n]^2 }, \\
d_{ra} &= \sqrt{ \|{w}_r - {w}_a\|^2 + z_r^2 }.
\end{aligned}
\end{equation}

\subsubsection{JU and JR Links}

We consider two interference links involving a ground-based jammer: the jammer-to-UAV (JU) and jammer-to-RIS (JR) links. These are modeled using Rician fading with elevation-angle-dependent $K$-factors to reflect the impact of spatial geometry.

JU Link is modeled as:
\begin{equation}
\begin{aligned}
\mathbf{h}_{ju}[n]= & \sqrt{\beta_0 d_{ju}^{-\alpha_{ju}}[n]} \\
                   & \left( \sqrt{\frac{K_{ju}[n]}{1 + K_{ju}[n]}} \, \mathbf{h}_{ju}^{\text{LoS}}[n] + \sqrt{\frac{1}{1 + K_{ju}[n]}} \, \mathbf{h}_{ju}^{\text{NLoS}}[n] \right),
\end{aligned}
\end{equation}
where $d_{ju}[n] = \sqrt{ \| {w}_u[n] - {w}_j \|^2 + z_u[n]^2 }$, $\alpha_{ju}$ is the path loss exponent, $\mathbf{h}_{ju}^{\text{LoS}}[n]$ is the deterministic LoS component (e.g., phase shift), $\mathbf{h}_{ju}^{\text{NLoS}}[n] \sim \mathcal{CN}(0,1)$ is the Rayleigh fading NLoS component, $K_{ju}[n]$ is an elevation-dependent Rician $K$-factor.

The elevation-based Rician $K$-factor is defined as:
\begin{equation}
K_{ju}[n] = \xi_1 \exp\left( \xi_2 \theta_{ju}[n] \right),
\end{equation}
where:
\begin{equation}
\theta_{ju}[n] = \arcsin\left( \frac{z_u[n] - z_j}{d_{ju}[n]} \right),
\end{equation}
is the elevation angle from jammer to UAV, and $\xi_1, \xi_2$ are fitting constants depending on the environment.

JR Link is modeled as:
\begin{equation}
\begin{aligned}
\mathbf{h}_{jr}[n] = &\sqrt{\beta_0 d_{jr}^{-\alpha_{jr}}[n]} \\
                     &\left( \sqrt{\frac{K_{jr}}{1 + K_{jr}}} \, \mathbf{h}_{jr}^{\text{LoS}}[n] + \sqrt{\frac{1}{1 + K_{jr}}} \, \mathbf{h}_{jr}^{\text{NLoS}}[n] \right),
\end{aligned}
\end{equation}
where $d_{jr}[n] = \sqrt{ \| {w}_r - {w}_j \|^2 + z_r^2 }$ is the jammer-to-RIS distance, $\alpha_{jr}$ is the path loss exponent for the JR link, $K_{jr}$ is the Rician $K$-factor (assumed static for fixed RIS), $\mathbf{h}_{jr}^{\text{LoS}}[n] \in \mathbb{C}^{M \times 1}$ is the RIS URA array response vector, $\mathbf{h}_{jr}^{\text{NLoS}}[n] \sim \mathcal{CN}(\mathbf{0}, \mathbb{I})$ is the i.i.d. Rayleigh fading vector.

The LoS component $\mathbf{h}_{jr}^{\text{LoS}}[n]$ is modeled as a 2D URA response vector:
\begin{equation}
\mathbf{h}_{jr}^{\text{LoS}}[n] = \mathbf{a}_{r}^{\text{JR}}[n] = \mathbf{a}_{ry}^{\text{JR}}[n] \otimes \mathbf{a}_{rz}^{\text{JR}}[n],
\end{equation}
where:
\begin{align*}
\mathbf{a}_{ry}^{\text{JR}}[n] = &[1, e^{-j \frac{2\pi}{\lambda} d_y \sin\theta_{jr}[n] \cos\phi_{jr}[n]}, \\
                                &\ldots, e^{-j \frac{2\pi}{\lambda} (M_y - 1) d_y \sin\theta_{jr}[n] \cos\phi_{jr}[n]} ]^T, \\
\mathbf{a}_{rz}^{\text{JR}}[n] = &[1, e^{-j \frac{2\pi}{\lambda} d_z \sin\theta_{jr}[n] \sin\phi_{jr}[n]}, \\
                                &\ldots, e^{-j \frac{2\pi}{\lambda} (M_z - 1) d_z \sin\theta_{jr}[n] \sin\phi_{jr}[n]} ]^T.
\end{align*}

The spatial angles $\theta_{jr}[n]$ and $\phi_{jr}[n]$ are calculated as:
\begin{equation}
\begin{aligned}
\sin \theta_{jr}[n] &= \frac{z_r - z_j}{d_{jr}[n]}, \\
\cos \phi_{jr}[n] &= \frac{y_r - y_j}{\sqrt{(x_r - x_j)^2 + (y_r - y_j)^2}}, \\
\sin \phi_{jr}[n] &= \frac{x_r - x_j}{\sqrt{(x_r - x_j)^2 + (y_r - y_j)^2}}.
\end{aligned}
\end{equation}

\subsection{Task Offloading}

To enable coordinated edge computing, each time slot of duration $\delta_t$ is divided into $\mathcal{S}$ sub-slots. The first $\mathcal{S}/2$ sub-slots are allocated for task offloading from UE $k \in \mathcal{K}$ to its associated UAV $u \in \mathcal{U}$, with $\tau_{u,k}^o[n]$ denoting the duration allocated for this transmission during slot $n$. The remaining $\mathcal{S}/2$ sub-slots are used for relaying data from UAV $u$ to the AP, with $\tau_{u,k}^R[n]$ denoting the relay duration. The total duration of both phases in each time slot must not exceed the slot length:
\begin{equation}
\sum_{u \in \mathcal{U}} \sum_{k \in \mathcal{K}} \left( \tau_{u,k}^o[n] + \tau_{u,k}^R[n] \right) \leq \delta_t, \quad \forall n.
\end{equation}

Each offloaded task is partitioned at the UAV: one part is computed locally, while the rest is forwarded to the AP. UAVs operate in a decode-and-forward (DF) relay mode with a one-slot delay. Therefore, slot $n=1$ is only used for receiving data from UEs, and slot $n=N$ is reserved for final computation, captured by:
$
\tau_{u,k}^R[0] = 0$, $ \tau_{u,k}^R[N] = 0$, $ \tau_{u,k}^o[N] = 0$, $ \forall u \in \mathcal{U}$, $k \in \mathcal{K}$.

Let $s_k[n]$ denote the transmitted symbol from UE $k$ at time slot $n$, modeled as a unit-variance symbol, i.e., $\mathbb{E}[|s_k[n]|^2] = 1$. Each UE transmits with power $p_t$. The received signal at UAV $u$ is:
\begin{equation}
\mathbf{y}_{u,k}[n] = \sqrt{p_t} \, \mathbf{h}_{u,k}[n] \, s_k[n] + \mathbf{n}_u[n],
\end{equation}
where $\mathbf{h}_{u,k}[n] \in \mathbb{C}^{L \times 1}$ is the channel from UE $k$ to UAV $u$, and $\mathbf{n}_u[n] \sim \mathcal{CN}(\mathbf{0}, \sigma^2 \mathbb{I}_L)$ is the AWGN at UAV $u$.

To decode $s_k[n]$, UAV $u$ applies a receive beamforming vector $\mathbf{u}_{u,k}[n] \in \mathbb{C}^{L \times 1}$, yielding:
\begin{equation}
\begin{aligned}
\hat{s}_{u,k}[n] &= \mathbf{u}_{u,k}^H[n] \, \mathbf{y}_{u,k}[n] \\
                 &= \sqrt{p_t} \, \mathbf{u}_{u,k}^H[n] \, \mathbf{h}_{u,k}[n] \, s_k[n] + \mathbf{u}_{u,k}^H[n] \, \mathbf{n}_u[n].
\end{aligned}
\end{equation}

The resulting SNR at UAV $u$ is given by:
\begin{equation}
\gamma_{u,k}[n] = \frac{p_t \left| \mathbf{u}_{u,k}^H[n] \mathbf{h}_{u,k}[n] \right|^2}{\sigma^2 \| \mathbf{u}_{u,k}[n] \|^2}.
\end{equation}

\subsubsection{Relay Phase to AP}

In the second half of each time slot, UAV $u$ forwards the decoded data $\tilde{s}_{u,k}[n]$ to the AP. The equivalent channel including both the direct UAV-AP link and the RIS-assisted reflection is denoted by:
\begin{equation}
\bar{\mathbf{h}}_{ua}^H[n] = \mathbf{h}_{ua}^H[n] + \mathbf{h}_{ra}^H[n] \mathbf{\Theta}[n] \mathbf{h}_{ur}[n],
\end{equation}
as previously defined. The transmitted signal is:
\begin{equation}
\mathbf{Y}_{u,k}[n] = \bar{\mathbf{h}}_{ua}^H[n] \, \mathbf{w}_{u,k}[n] \, \tilde{s}_{u,k}[n] + n_a[n],
\end{equation}
where $\mathbf{w}_{u,k}[n] \in \mathbb{C}^{L \times 1}$ is the transmit beamforming vector of UAV $u$ and $n_a[n] \sim \mathcal{CN}(0, \sigma^2)$ is the noise at the AP.

The received SNR at the AP is:
\begin{equation}
\gamma_{a,u,k}[n] = \frac{\left| \bar{\mathbf{h}}_{ua}^H[n] \, \mathbf{w}_{u,k}[n] \right|^2}{\sigma^2 + \mathcal{I}_{u,k}[n]},
\end{equation}
where $\mathcal{I}_{u,k}[n]$ denotes the interference power at the AP (e.g., due to jammer), which can be modeled separately depending on the jammer’s strategy. For example:
\[
\mathcal{I}_{u,k}[n] = p_j \left( \| \mathbf{h}_{jr}[n] \|^2 + \| \mathbf{h}_{ju}[n] \|^2 \right),
\]
if both JR and JU links contribute to AP interference.

\subsubsection{Offloading Rate Model}

Let $B$ be the system bandwidth. The number of task bits offloaded from UE $k$ to UAV $u$ in time slot $n$ is:
\begin{equation}
l_{u,k}^o[n] = \tau_{u,k}^o[n] \, B \log_2\left(1 + \gamma_{u,k}[n]\right),
\end{equation}
while the number of task bits forwarded from UAV $u$ to the AP is:
\begin{equation}
l_{a,u,k}^R[n] = \tau_{u,k}^R[n] \, B \log_2\left(1 + \gamma_{a,u,k}[n]\right).
\end{equation}

\subsubsection{Computation Model}

Let $l_{u,k}^{\text{comp}}[n]$ be the number of task bits from UE $k$ processed by UAV $u$ in slot $n$. The onboard computing capacity at UAV $u$ is constrained by:
\begin{equation}
\sum_{k \in \mathcal{K}} l_{u,k}^{\text{comp}}[n] \, c_u \leq F_u^{\max} \delta_t, \quad \forall u \in \mathcal{U},
\end{equation}
where $c_u$ is the number of CPU cycles per bit, and $F_u^{\max}$ is the CPU capacity of UAV $u$ in cycles per second.

Similarly, the AP computation capacity is:
\begin{equation}
\sum_{u \in \mathcal{U}} \sum_{k \in \mathcal{K}} l_{a,u,k}^R[n] \, c_a \leq F_a^{\max} \delta_t.
\end{equation}

\subsubsection{Data Causality Constraint}

Due to the one-slot relay delay, the bits received by UAV $u$ at slot $n$ can only be computed or forwarded at slot $n+1$:
\begin{equation}
\begin{aligned}
l_{u,k}^o[n] &\leq l_{a,u,k}^R[n+1] + l_{u,k}^{\text{comp}}[n+1], \\
             &  \forall u \in \mathcal{U}, \, k \in \mathcal{K}, \, n = 1, \ldots, N-1.
\end{aligned}
\end{equation}

\subsubsection{UE Local Computation and Task Completion}

Let $l_{u,k}^{\text{loc}}[n]$ be the number of bits computed locally at UE $k$ during time slot $n$. The total number of processed bits by UE $k$ in slot $n$ is:
\begin{equation}
L_k[n] = l_{u,k}^{\text{loc}}[n] + l_{u,k}^o[n].
\end{equation}

To meet application-level processing requirements, each UE must process at least $L_k$ bits over the total horizon:
\begin{equation}
\sum_{n=1}^N \left( l_{u,k}^{\text{loc}}[n] + l_{u,k}^o[n] \right) \geq L_k,  \forall k \in \mathcal{K}.
\end{equation}

\subsection{Energy Consumption}

The system energy consumption includes communication, computation, and propulsion energy from both UEs and UAVs. We model these components as follows:

\subsubsection{ UE Energy Consumption}

Each UE $k \in \mathcal{K}$ incurs energy in two ways: offloading to a UAV and local task processing.
\begin{enumerate}[label=\roman*.]
    \item {Transmission energy}: The energy consumed by UE $k$ to offload data to its associated UAV $u$ in time slot $n$ is:
    \begin{equation}
    E_{k}^{\text{tx}}[n] = p_{u,k}[n] \, \tau_{u,k}^o[n],
    \end{equation}
    where $p_{u,k}[n]$ is the transmission power of UE $k$ during slot $n$.
    \item {Local computation energy}: Modeled using dynamic voltage and frequency scaling (DVFS) theory \cite{wu2025towards}, the energy consumed by UE $k$ for local processing is:
    \begin{equation}
    E_{k}^{\text{comp}}[n] = \frac{\mu_{k} \left( l_{u,k}^{\text{loc}}[n] \right)^3}{\delta_t^2},
    \end{equation}
    where $\mu_{k}$ is the switched capacitance coefficient of UE $k$, and $l_{u,k}^{\text{loc}}[n]$ is the number of locally processed bits.
\end{enumerate}

\subsubsection{ UAV Energy Consumption}
Each UAV $u \in \mathcal{U}$ consumes energy for computation and flying.
\begin{enumerate}[label=\roman*.]
    \item {Computation energy}: To process offloaded bits from all associated UEs, UAV $u$ consumes:
    \begin{equation}
    E_u^{\text{comp}}[n] = \sum_{k \in \mathcal{K}} \frac{\mu_u \left( l_{u,k}^{\text{comp}}[n] \right)^3}{\delta_t^2},
    \end{equation}
    where $\mu_u$ is the capacitance coefficient of UAV $u$.
    \item {Propulsion energy}: The flying energy of UAV $u$ at slot $n$ is modeled as:
    \begin{equation}
    E_u^{\text{fly}}[n] = \delta_t \left( \vartheta_1 \left\| \mathbf{V}_u[n] \right\| + \frac{\vartheta_2}{\left\| \mathbf{V}_u[n] \right\|} \right),
    \end{equation}
    where 
        \begin{equation}
            \mathbf{V}_u[n] = [V_{u,x}[n], V_{u,y}[n], V_{u,z}[n]]^T
        \end{equation}
         is the UAV velocity vector,
             \begin{equation}
             \left\| \mathbf{V}_u[n] \right\| = \sqrt{ V_{u,x}^2[n] + V_{u,y}^2[n] + V_{u,z}^2[n] }
            \end{equation}
         is the UAV’s speed,
         $\vartheta_1, \vartheta_2$ are aerodynamic constants reflecting linear and inverse-speed power terms (e.g., air drag and hover-related inefficiencies).
\end{enumerate}

\subsubsection{ Total Energy Consumption}
The total system energy consumption at time slot $n$ is the sum of all UE and UAV contributions:
\begin{equation}
E[n] = \sum_{k \in \mathcal{K}} \left( E_k^{\text{tx}}[n] + E_k^{\text{comp}}[n] \right) + \sum_{u \in \mathcal{U}} \left( E_u^{\text{comp}}[n] + E_u^{\text{fly}}[n] \right).
\end{equation}

\section{{Problem Formulation}}
We aim to maximize the system-wide energy efficiency (in bits/Joule) over the total $N$ time slots:
{
\begin{equation}
\begin{aligned}
\max_{z}
\frac{
\sum\limits_{n=1}^{N}\sum\limits_{k\in\mathcal{K}}
\Big( l_{u,k}^{\mathrm{loc}}[n] + l_{u,k}^{o}[n] \Big)
}{
\sum\limits_{n=1}^{N}
\left[
\begin{aligned}
&\sum\limits_{k\in\mathcal{K}}
\Big( E_{u,k}^{\mathrm{tx}}[n] + E_{u,k}^{\mathrm{comp}}[n] \Big) \\
&\quad + \sum\limits_{u\in\mathcal{U}}
\Big( E_{u}^{\mathrm{comp}}[n] + E_{u}^{\mathrm{fly}}[n] \Big)
\end{aligned}
\right]
}
\end{aligned}
\end{equation}
}

Subject to the following constraints:

{[UAV Kinematics Constraints]}
\begin{equation}
\begin{aligned}
\text{C1:} & \quad \mathbf{q}_u[n+1] = \mathbf{q}_u[n] + \mathbf{V}_u[n] \delta_t + \tfrac{1}{2} \mathbf{A}_u[n] \delta_t^2, \quad \forall u,\, n, \\
\text{C2a:} & \quad \mathbf{q}_u[0] = \mathbf{q}_{u,0}, \quad \mathbf{q}_u[N] = \mathbf{q}_{u,F}, \quad \forall u, \\
{\text{C2b:}} & \quad {
\begin{cases}
x_{\min} \leq x_{u}[n] \leq x_{\max}, \\
y_{\min} \leq y_{u}[n] \leq y_{\max}, \\
z_{\min} \leq z_{u}[n] \leq z_{\max},
\end{cases}
 \forall u,\, n, }\\
\text{C3:} & \quad \mathbf{V}_u[n+1] = \mathbf{V}_u[n] + \mathbf{A}_u[n] \delta_t, \quad \forall u,\, n, \\
\text{C4a:} & \quad \|\mathbf{q}_u[n] - \mathbf{q}_u[n-1]\| \leq \delta_t V_{\max}, \quad \forall u,\, n = 2, \dots, N, \\
{\text{C4b:}} & \quad {\left\|\mathbf{q}_u[n]-\mathbf{q}_{u, F}\right\|_2 \leq(N-n) V_{\max } \delta_t }\\
\text{C5:} & \quad \|\mathbf{V}_u[n]\| \leq V_{\max}, \quad \forall u,\, n.
\end{aligned}
\end{equation}
{where C2b enforces spatial boundary conditions for UAV flight, ensuring that each UAV remains within the operational area $(x_{\min},x_{\max})\times(y_{\min},y_{\max})$ and within the altitude range $[z_{\min},z_{\max}]$. C4b ensures that from any time step, the UAV can still reach the destination within the remaining time slots under the maximum feasible velocity constraint.}

{[RIS Phase Constraints]}
\begin{equation}
\begin{aligned}
\text{C6:} & \quad \theta_m[n] \in [0, 2\pi), \quad \forall m = 1, \dots, M, \forall n, \\
\text{C7:} & \quad \mathbf{\Theta}[n] = \operatorname{diag}(e^{j \theta_1[n]}, \dots, e^{j \theta_M[n]}) \in \mathbb{C}^{M \times M}, \quad \forall n.
\end{aligned}
\end{equation}

{[Time Allocation Constraints]}
\begin{equation}
\begin{aligned}
\text{C8:} & \quad \tau_{u,k}^o[n] \geq 0, \quad \tau_{u,k}^R[n] \geq 0, \quad \forall u,\, k,\, n, \\
\text{C9:} & \quad \tau_{u,k}^R[0] = 0, \quad \tau_{u,k}^R[N] = 0, \quad \tau_{u,k}^o[N] = 0, \quad \forall u,\, k, \\
\text{C10:} & \quad \sum_{u \in \mathcal{U}} \sum_{k \in \mathcal{K}} \left( \tau_{u,k}^o[n] + \tau_{u,k}^R[n] \right) \leq \delta_t, \quad \forall n.
\end{aligned}
\end{equation}

{[UAV Transmission Power Constraints]}
\begin{equation}
\begin{aligned}
\text{C11:} & \quad \| \mathbf{w}_{u,k}[n] \|^2 \leq P_u^{\max}, \quad \forall u,\, k,\, n, \\
\text{C12:} & \quad \frac{1}{N \delta_t} \sum_{n=1}^N \sum_{k \in \mathcal{K}} \tau_{u,k}^R[n] \, \| \mathbf{w}_{u,k}[n] \|^2 \leq P_u^{\text{ave}}, \quad \forall u.
\end{aligned}
\end{equation}

{[Task and Computation Constraints]}
\begin{equation}
\begin{aligned}
\text{C13:} & \quad l_{u,k}^o[n] \leq l_{a,u,k}^R[n{+}1] + l_{u,k}^{\text{comp}}[n{+}1],  \quad \forall u,k,\, n, \\
\text{C14:} & \quad \sum_{k \in \mathcal{K}} l_{u,k}^{\text{comp}}[n] c_u \leq F_u^{\max} \delta_t, \quad \forall u,\, n, \\
\text{C15:} & \quad \sum_{u \in \mathcal{U}} \sum_{k \in \mathcal{K}} l_{a,u,k}^R[n] c_a \leq F_a^{\max} \delta_t, \quad \forall n, \\
\text{C16:} & \quad \sum_{n=1}^N \left( l_{u,k}^{\text{loc}}[n] + l_{u,k}^o[n] \right) \geq L_k, \quad \forall k.
\end{aligned}
\end{equation}

\vspace{1em}
{Optimization variables:}
\begin{align*}
z = \big\{ 
 \mathbf{V}_u[n], \mathbf{A}_u[n], 
\mathbf{\Theta}[n],
 \tau_{u,k}^o[n], \tau_{u,k}^R[n]\big\}.
\end{align*}

\begin{figure*}[htbp]
  \centering
  \includegraphics[width=\textwidth]{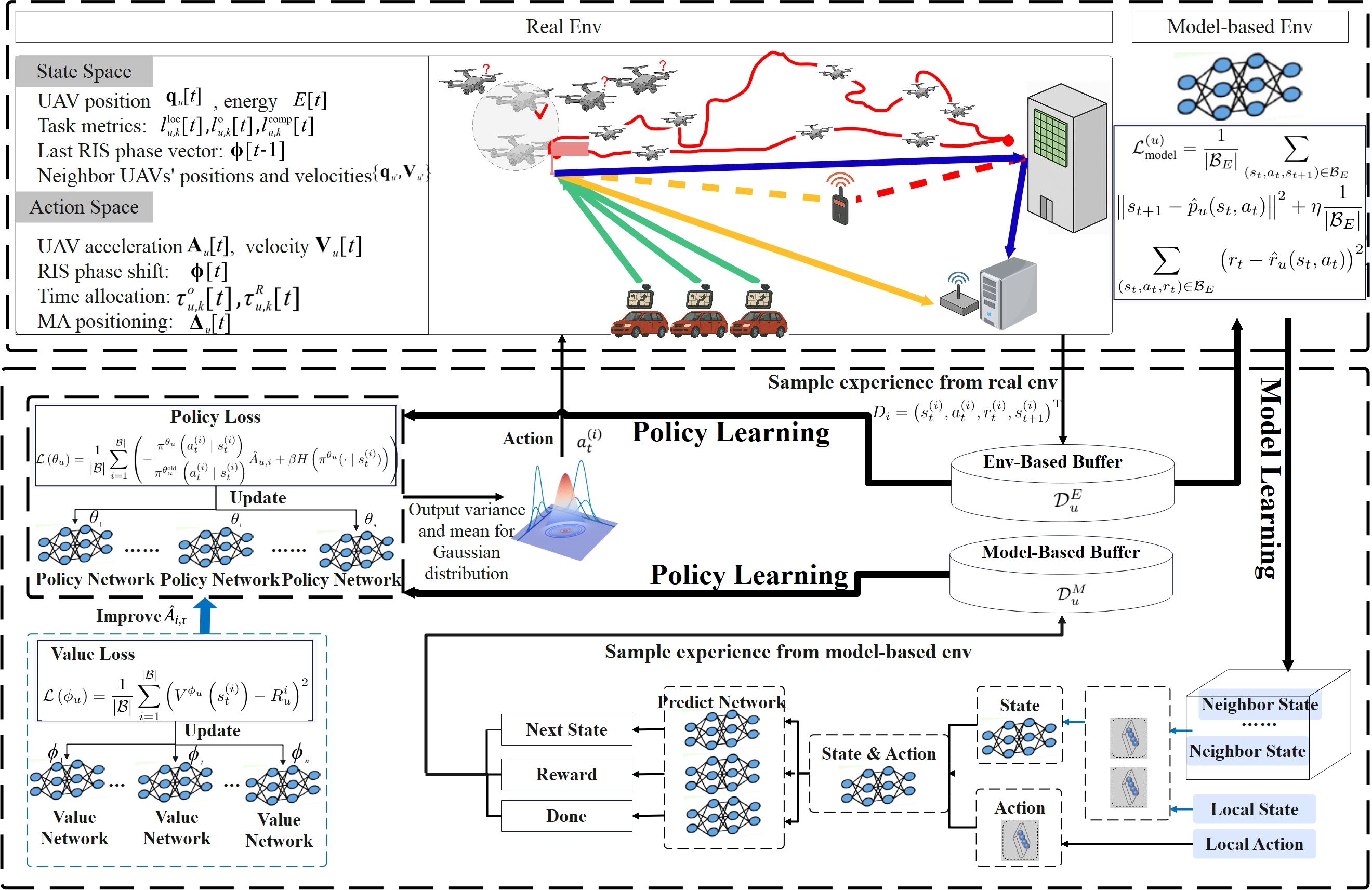}
  \caption{Workflow of the decentralized model-based RL for UAV–RIS–MEC systems. The diagram illustrates how real-environment and model-based replay buffers are leveraged to train the policy (actor) and value (critic) networks via $T$-step branched rollouts with local and $\kappa$-hop neighbor observations, and how PPO-style updates refine both networks.}
  \label{fig:rl_architecture}
\end{figure*}

\begin{figure*}[htbp]
  \centering
  \includegraphics[width=\textwidth]{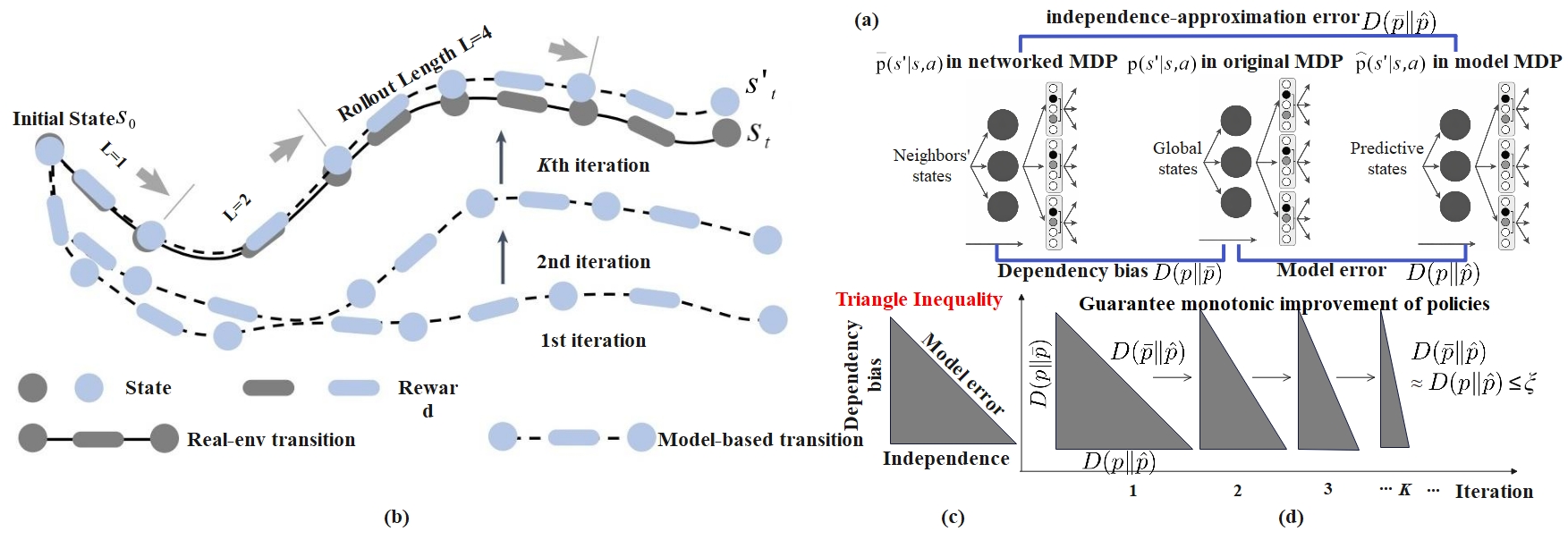}
    \caption{In our approach: {(a) We distinguish three transition kernels: (i) $p(s'|s,a)$, the true environment dynamics; (ii) $\bar{p}(s'|s,a)$, the factorized ``networked'' dynamics using only local $\kappa$-hop information (induces dependency bias); (iii) $\hat{p}(s'|s,a)$, the learned predictive model (incurs model error). Their discrepancies give rise to dependency bias $D(p\|\bar{p})$, model error $D(p\|\hat{p})$, and independence-approximation error $D(\bar{p}\|\hat{p})$.}  (b) The model learning process repeatedly samples {experiences} from the model buffer. Here, $s_t$ denotes the actual system state at step $t$, while $s'_t$ indicates the state predicted by the model at the same step. (c) With short branched rollouts, model-predicted transitions align with real dynamics and $D(p \| \hat{p}) \approx D(p \| \bar{p}) \le \xi$, {where $\xi$ is the upper bound of model error tolerance.} This ensures a monotonic policy improvement.}
  \label{fig:triangular}
\end{figure*}

\section{Model-Based Decentralized RL}

To address joint trajectory planning, RIS control, and task offloading under limited global observability and low sample efficiency, we propose a model-based decentralized reinforcement learning (MB-DRL) framework for multi-UAV, multi-UE RIS-assisted MEC networks. Each UAV (agent) $u \in \mathcal{U}$ learns a policy {$\pi^{\theta_u}$} from its local state and $\kappa$-hop neighbor interactions{, without requiring global observability.}

\subsection{Motivation}

Our method targets two coupled issues. First, global information is limited: distributed UAVs operate with sparse coordination, so each agent only observes its own state and those of $\kappa$-hop neighbors $\mathcal{N}_u^{(\kappa)}$. Second, model-free exploration is sample-inefficient: real-world data are costly, {while purely analytical link and mobility models miss uncertainties (interference, blockage, nonstationary traffic).} We therefore let each agent learns a localized predictive model for short-horizon rollouts that enhance sample efficiency and stabilize PPO-based policy updates under partial observability.

\subsubsection{Why neighbor UAV states matter?}

Decisions on trajectory, RIS phases, and task allocation are coupled through wireless and kinematic constraints. From {(\ref{eq:snr_ap})}, the received signal at the AP depends on the RIS phase matrix $\mathbf\Theta[n]$, which is jointly affected by concurrent agent proposals and can cause beam conflicts and interference. Constraint C10 implies a {shared per-slot time budget}: $\tau_{u,k}^o[n]$ and $\tau_{u,k}^R[n]$. Kinematic constraints C1–C4 imply path overlap increases collision risk and propulsion energy, while task variables $l_{u,k}^o[n]$, $l_{u,k}^{\text{comp}}[n]$ depend on SNR and candidate agents’ loads, both shaped by neighbors’ locations and queues. Hence each agent observes $\kappa$-hop neighbors’ positions, velocities, queues, and recent RIS controls to make decentralized yet coordinated decisions.

\subsubsection{Comparison with traditional MARL}
CTDE methods such as MADDPG assume global state access and a centralized critic, which degrades scalability under partial observability and constrained links. Our approach is fully decentralized: each agent uses only local and $\kappa$-hop observations, learns a predictive transition model, and employs short-horizon model-based rollouts for efficiency. We adopt PPO with parameterized policy $\pi^{\theta_u}$ and value $V^{\phi_u}$ because the clipped surrogate yields stable, near-monotonic improvement under noisy rewards and high-dimensional states, and supports parallel local training.

\subsection{{MDP Formulation}}
Each agent $u$ acts on a local state containing its 3D position $\mathbf{q}_u[n]$ and residual energy; per-user task metrics $l_{u,k}^{\text{loc}}[n]$, $l_{u,k}^{o}[n]$, and $l_{u,k}^{\text{comp}}[n]$; the last RIS phase vector $\boldsymbol{\Phi}[n-1]=\{\phi_m[n-1]\}_{m=1}^{M}$, and the corresponding diagonal phase-shift matrix $\boldsymbol{\Theta}[n-1]=\mathrm{diag}(e^{\mathrm{j}\phi_1[n-1]},\ldots,e^{\mathrm{j}\phi_M[n-1]})$; as well as the positions and velocities of neighbors in $\mathcal{N}_u^{(\kappa)}$:$\left\{ \mathbf{q}_{u'}, \mathbf{V}_{u'} \right\}_{u' \in \mathcal{N}_u^{(\kappa)}}$. The action consists of UAV acceleration $\mathbf{A}_u[n]$ (and the implied velocity), time allocations $\tau_{u,k}^{o}[n]$ and $\tau_{u,k}^{R}[n]$, and a local RIS phase proposal $\boldsymbol{\Phi}_u[n]$. {A lightweight controller co-located with the RIS aggregates proposals as $\boldsymbol{\Phi}[n] = \mathrm{Agg}\big(\{\boldsymbol{\Phi}_u[n]\}_{u\in\mathcal{U}}\big)$, where $\mathrm{Agg}(\cdot)$ is a stateless function using only the submitted proposals (e.g., weighted averaging:$\mathbf{\Theta}[n]=\bar{\mathbf\Theta}[n]=\frac{\sum_{u=1}^U \mathbf\Theta_u[n]}{U}$, $\boldsymbol{\Theta}_u[n]=\operatorname{diag}(\boldsymbol{\Phi}_u[n])=\operatorname{diag}\left(e^{j \theta_{u,1}[n]}, e^{j \theta_{u,2}[n]}, \ldots, e^{j \theta_{u,M}[n]}\right)$, or winner-takes-most on a codebook). }

The per-slot reward for agent $u$ is as:
\begin{equation}
\begin{aligned}
r_u[n] &= \frac{\sum_{k \in \mathcal{K}} \big( l_{u,k}^{\text{loc}}[n] + l_{u,k}^{o}[n] \big)}
{\sum_{k \in \mathcal{K}} \big( E_{u,k}^{\text{tx}}[n] + E_{u,k}^{\text{comp}}[n] \big)
+ E_u^{\text{fly}}[n] }
- \beta I_u[n],
\end{aligned}
\label{eq:reward}
\end{equation}
where $I_u[n]$ measures interference imposed on neighbors via overlapping RIS beams or transmit directions; here $\beta$ controls the throughput–contention tradeoff {(the same symbol $\beta$ also appears later as the entropy regularization weight in the PPO loss; meanings are distinguished by context).}

\subsection{{Localized Communication}}
{Define the $\kappa$-hop neighborhood of UAV $u$ as the set of UAVs that are geographically closest to $u$ at time slot $n$ :
\begin{equation}
\mathcal{N}_u^{(\kappa)}[n]=\underset{\substack{\mathcal{S} \subseteq \mathcal{U} \backslash\{u\},|\mathcal{S}|=\kappa}}{\arg \min } \sum_{v \in \mathcal{S}}\left\|\mathbf{q}_u[n]-\mathbf{q}_v[n]\right\|_2
\end{equation}
where $\mathbf{q}_u[n] \in \mathbb{R}^3$ denotes the 3D position. }

{Unlike I3CNet \cite{singh2018learning}, where each UAV $u$ updates its hidden state by simple aggregation
\begin{equation}
\begin{aligned}
h_u[n] = \operatorname{LSTM}\big(h_u[n-1], \operatorname{relu}(s_{\mathcal{N}_u^{(\kappa)}}[n])\big),
\end{aligned}
\label{eq:i3cnet}
\end{equation}
which fails to capture nonlinear coupled dynamics; our method introduces localized inter-agent communication through explicit neighborhood aggregation. Specifically,  each UAV $u$ updates its internal state by combining its own processed input $\operatorname{relu}(s_{\mathcal{N}_u^{(\kappa)}}[n])$ and the recent behaviors (policy $\pi_u[n{-}1]$ and hidden state $h_u[n{-}1]$):
\begin{equation}
\begin{aligned}
h_u[n]
=\operatorname{LSTM}\Big(
h_u[n{-}1],
\operatorname{concat}\big(
&\operatorname{relu}(s_{\mathcal{N}_u^{(\kappa)}}[n]),\\
&\operatorname{relu}(\pi_{u}[n-1]),\\
&\operatorname{relu}(h_u[n-1])
\big)
\Big).
\end{aligned}
\label{eq:ours}
\end{equation}
This formulation realizes local communication via direct neighborhood aggregation—each agent exchanges only low-dimensional hidden and policy features within its $\kappa$-hop neighborhood, without any global broadcast.}

{For comparison, distributed PPO (DPPO) and centralized PPO (CPPO) represent two opposite extremes of communication scope:
\begin{itemize}
\item In CPPO, a centralized critic or coordinator collects all agents’ states and actions for joint updates:
\begin{equation}
V[n]=\operatorname{LSTM}\left(\operatorname{concat}\left(\left\{s_u[n], a_u[n]\right\}_{u=1}^N\right)\right),
\end{equation}
This achieves full observability but suffers from heavy communication and poor scalability.
\item In DPPO, each agent learns completely independently, using only its local trajectory data:
\begin{equation}
V_u[n]=\operatorname{LSTM}(\operatorname{relu}\left(s_u[n]\right)),
\end{equation}
which eliminates communication but cannot model inter-agent coupling.
\end{itemize}
By contrast, our proposed neighbor-aggregated update provides a middle ground between CPPO’s global coupling and DPPO’s isolation.}

{Although each UAV observes only its $\kappa$-hop neighborhood, the approximation error introduced by ignoring distant agents is theoretically bounded. The formal notion of a $\xi$-dependent networked system (Appendix~B) shows that if the inter-agent coupling decays sufficiently fast, the global dynamics can be well approximated by local transition models within a bounded error $\xi$.}

\subsection{Policy Learning}
{Note that all “true dynamics” discussed in this paper are instantiated by the system level model. The training objective of the model-based environment is therefore twofold: to approximate the system model while being corrected by real reward feedback. In other words, the learned model is not merely a surrogate to replicate the analytical system model, but rather a predictive environment aligned with maximizing task rewards. Since the system model cannot fully capture real-world uncertainties, reward signals are indispensable to ground the learning process.}

Each agent $u$ maintains a policy $\pi^{\theta_u}$, a value function $V^{\phi_u}$, a local transition model $\hat{p}_u$, and two FIFO replay buffers: an environment buffer $\mathcal{D}^E_u$ (capacity $N_E$) and a model buffer $\mathcal{D}^M_u$ (capacity $N_M$). A single sampled transition (mini-batch index $i$) is written as
\begin{equation}
D_i = \big(s_t^{(i)}, a_t^{(i)}, r_t^{(i)}, s_{t+1}^{(i)}\big)^{\mathrm{T}},
\end{equation}
from a mini-batch $\mathcal{B} = (D_1, D_2, \ldots, D_{|\mathcal{B}|})$, where $s_t^{(i)}$ denotes the observed state at time $t$ for the $i$-th sample. Each $D_i$ is drawn from the union buffer: $D_i \in \mathcal{D}^E_u \cup \mathcal{D}^M_u$.

Training iterates over four phases. (i) {Real interaction.} Execute $a_u[n] \sim \pi^{\theta_u}(\cdot \mid s_t)$, observe $r_u[n]$ and the next state $s_{t+1}$, and push $(s_t, a_u[n], r_u[n], s_{t+1})$ into $\mathcal{D}^E_u$. (ii) {Environment-model learning.} Fit the one-step predictor $\hat{p}_u$ with supervised regression on minibatches $\mathcal{B}_E \subset \mathcal{D}^E_u$ as Eq.~(\ref{eq:model}), where the second term is optional (reward head) with weight $\eta\ge0$. (iii) {Branched rollouts.} Sample anchor states from $\mathcal{D}^E_u$, roll out $\hat{p}_u$ for a short horizon $T$ under the current policy to generate $(\hat{s}_{t+1}, a_{t+1}, \hat{r}_{t+1}, \hat{s}_{t+2}),\ldots$, and push them into $\mathcal{D}^M_u$.  (iv) Policy improvement. Form mixed mini-batches $\mathcal{B}\subset \mathcal{D}^E_u \cup \mathcal{D}^M_u$, compute advantages $\hat{A}_{u,i}$ (e.g., generalized advantage estimator (GAE) with parameter $\lambda_{\mathrm{GAE}}\in[0,1]$): {for each sampled time index $t$ in mini-batch $i$, define the TD residual
\begin{equation}
\delta^{(i)}_{u,t} = r^{(i)}_{t}
+ \gamma\,(1-d^{(i)}_{t})\,V^{\phi_u}\!\left(s^{(i)}_{t+1}\right)
- V^{\phi_u}\!\left(s^{(i)}_{t}\right),
\label{eq:tdres}
\end{equation}
where $d^{(i)}_{t}\in\{0,1\}$ indicates trajectory termination (or branch end) at $t$.
\begin{equation}
\hat{A}^{(i)}_{u,t}
= \sum_{\ell=0}^{L^{(i)}_t-1}
\big(\gamma \lambda_{\mathrm{GAE}}\big)^{\ell}
\!\!\left(\prod_{j=0}^{\ell-1} (1-d^{(i)}_{t+j})\right)
\delta^{(i)}_{u,t+\ell},
\label{eq:gae}
\end{equation}
with truncation length $L^{(i)}_t$ given by the remaining real/model steps until termination or branch end.
The bootstrapped return target is
\begin{equation}
R^{(i)}_{u,t} = \hat{A}^{(i)}_{u,t} + V^{\phi_u}\!\left(s^{(i)}_{t}\right).
\label{eq:return}
\end{equation}}
Update parameters with Eq.~(\ref{eq:policy}) and Eq.~(\ref{eq:value}).

{The theoretical basis of this model-based learning strategy is established in Appendix A, which proves that the true return $\eta[\pi]$ is lower-bounded by the model-estimated return $\hat{\eta}[\pi]$ minus a bounded discrepancy term $C\left(p, \hat{p}, \pi, \pi_{\mathrm{D}}\right)$. This guarantee ensures safe and consistent policy improvement, as long as each update increases the model-based return more than the possible modeling error-precisely the condition maintained in our algorithm through frequent model retraining and PPO clipping.}

\begin{figure*}{
\begin{equation}
\mathcal{L}^{(u)}_{\text{model}}
=\frac{1}{|\mathcal{B}_E|}\sum_{(s_t,a_t,s_{t+1})\in\mathcal{B}_E}
\big\| s_{t+1}-\hat{p}_u(s_t,a_t) \big\|^2
+\eta\frac{1}{|\mathcal{B}_E|}\sum_{(s_t,a_t,r_t)\in\mathcal{B}_E}
\big(r_t-\hat{r}_u(s_t,a_t)\big)^2,
\label{eq:model}
\end{equation}
\begin{equation}
\mathcal{L}\left(\theta_u\right)
=\frac{1}{|\mathcal{B}|} \sum_{i=1}^{|\mathcal{B}|}\left(
-\frac{\pi^{\theta_u}\left(a_t^{(i)} \mid s_t^{(i)}\right)}{\pi^{\theta_u^{\text {old }}}\left(a_t^{(i)} \mid s_t^{(i)}\right)} \hat{A}_{u,i}
+\beta H\left(\pi^{\theta_u}(\cdot\mid s_t^{(i)})\right)\right),
\label{eq:policy}
\end{equation}
\begin{equation}
\mathcal{L}\left(\phi_u\right)
=\frac{1}{|\mathcal{B}|} \sum_{i=1}^{|\mathcal{B}|}\left(V^{\phi_u}\left(s_t^{(i)}\right)-R_{u,t}^{(i)}\right)^2.
\label{eq:value}
\end{equation}
\noindent\rule{\textwidth}{0.4pt}}
\end{figure*}
\subsection{Branched Rollout and Model Error Control}
From each real transition in $\mathcal{D}^E_u$, the model simulates only the next $T{-}1$ steps. Fig.~\ref{fig:triangular} visualizes the process: panel (a) contrasts predicted and real trajectories; panel (b) shows model error constrained; panel (c) highlights steady policy improvement. We track a bounded tolerance $\xi$ via $D(p\parallel \hat{p})\approx D(p\parallel \bar{p})=\xi$, but implementation relies on the operational criterion above (short $T$, anchor-at-real states).

As formally analyzed in Appendix C, long-horizon (vanilla) model rollouts can accumulate model bias and lead to loose performance guarantees, while short $T$-step branched rollouts effectively bound the discrepancy between true and modeled returns.
The derived bound (Eq. (66)) shows that the influence of model error $\epsilon_{m_i}$ is truncated after $T$ steps, which explains the empirically observed monotonic improvement in Fig.~\ref{fig:triangular}(d). Furthermore, Corollary 1 extends this guarantee to the general $\xi$-dependent case, confirming that even with limited inter-agent coupling, the expected return deviation remains bounded and diminishes.

{The step-by-step training procedure of the proposed MB-DRL framework is summarized in Algorithm~\ref{alg:mbdrl_short}.}

\begin{algorithm}[t]
\caption{{Model-Based Decentralized RL  (MB-DRL)}}
\label{alg:mbdrl_short}
{\color{black}
\begin{algorithmic}[1]
\REQUIRE Agents $\mathcal{U}$, horizon $N$, rollout length $T$, neighborhood size $\kappa$
\STATE Initialize $\{\pi^{\theta_u},V^{\phi_u},\hat{p}_u, \mathcal{D}^E_u,\mathcal{D}^M_u\}_{u\in\mathcal{U}}$
\FOR{each training episode}
    \STATE \textbf{(I) Sampling}
    \STATE Reset environment
    \FOR{$n=0$ \TO $N-1$}
        \FOR{each UAV $u\in\mathcal{U}$}
            \STATE Observe $s_u[n]$ (local + $\kappa$-hop neighbors)
            \STATE Sample $a_u[n]\!\sim\!\pi^{\theta_u}(\cdot|s_u[n])$ and propose RIS phases $\boldsymbol{\Phi}_u[n]$
        \ENDFOR
        \STATE Aggregate RIS: $\boldsymbol{\Phi}[n]=\mathrm{Agg}(\{\boldsymbol{\Phi}_u[n]\}_{u\in\mathcal{U}})$ and apply $\mathbf{\Theta}[n]$
        \STATE Step environment; obtain $\{r_u[n], s_u[n{+}1]\}_{u\in\mathcal{U}}$
        \FOR{each UAV $u\in\mathcal{U}$}
            \STATE Store $(s_u[n],a_u[n],r_u[n],s_u[n{+}1])$ into $\mathcal{D}^E_u$
        \ENDFOR
    \ENDFOR

    \FOR{each UAV $u\in\mathcal{U}$}
        \STATE \textbf{(II) Local model learning}
        \STATE Sample minibatch $\mathcal{B}_E\subset \mathcal{D}^E_u$ and update $\hat{p}_u$ (and $\hat{r}_u$) via (\ref{eq:model})

        \STATE \textbf{(III) Branched rollout }
        \STATE Sample anchor states $\{\hat{s}_0\}$ from $\mathcal{D}^E_u$
        \FOR{$t=0$ \TO $T-1$}
            \STATE Sample $\hat{a}_t\!\sim\!\pi^{\theta_u}(\cdot|\hat{s}_t)$, predict $\hat{s}_{t+1}=\hat{p}_u(\hat{s}_t,\hat{a}_t)$ (and $\hat{r}_t$)
            \STATE Store $(\hat{s}_t,\hat{a}_t,\hat{r}_t,\hat{s}_{t+1})$ into $\mathcal{D}^M_u$
        \ENDFOR

        \STATE \textbf{(IV) PPO policy and value update}
        \STATE Form mixed minibatch $\mathcal{B}\subset \mathcal{D}^E_u\cup\mathcal{D}^M_u$
        \STATE Update $\theta_u$ and $\phi_u$ using (\ref{eq:policy})--(\ref{eq:value}) with GAE advantages
    \ENDFOR
\ENDFOR
\end{algorithmic}
}
\end{algorithm}

\section{Simulation}
\subsection{Setup}
\subsubsection{Network Topology and Mobility}
We consider a $1,000 \mathrm{m} \times 1,000 \mathrm{m}$ square area of interest (AoI). Three single-antenna UEs ($K=10$) are randomly distributed within a $200 \mathrm{m} \times 200 \mathrm{m}$ square centered at $(200,200)$ on the ground plane ($z=0$). A MEC-enabled AP with a single antenna is deployed at coordinates $(x_a, y_a)=(100,850)$ at ground level ($z_a=0$). A uniform rectangular RIS is placed at $(x_r, y_r)=(650,200)$ at height $z_r=20$ m, consisting of $M_y=8$ and $M_z=8$ elements, i.e., $M=64$ elements. The network is assisted by  {$U=10$} UAVs, each equipped with $L=4$ antennas, whose initial and final positions are $(0,0,20)$ and $(1000,800,20)$. {The altitude of each UAV is restricted within $z_{\min}=10$m and $z_{\max}=200$m, 
and must remain inside the AoI boundary.} In addition, a ground-based jammer is located at $(100,800)$ with fixed transmit power $p_j=0.5\mathrm{W}$.

\subsubsection{Time and Resource Allocation}
The total mission duration is $T=100\mathrm{s}$, uniformly divided into $N=100$ time slots of $\delta_t=1\mathrm{s}$ each. Every slot is further partitioned into six equal sub-slots. The available bandwidth is $B=1\mathrm{MHz}$. Each UE is assigned a task load of $L_k = 2\times10^5$ bits to be processed before the mission ends. The maximum computational capacities are $F_u^{\max}=10^9$ cycles/s for each UAV and $F_a^{\max}=10^{10}$ cycles/s for the AP, with per-bit computational complexity $c_u=c_a=10^3$ cycles/bit.

\subsubsection{Energy and Mobility Models}
DVFS parameters are set as $\mu_u = 10^{-28}$ and $\mu_{u,k}=10^{-28}$. UAV flight energy consumption is modeled using parameters $\vartheta_1 = 9.26$ and $\vartheta_2 = 2250$. UAV mobility is constrained by a maximum velocity $V_{\max}=20\mathrm{m/s}$ and an acceleration bound $|\mathbf{A}_u[n]|\leq 2\mathrm{m/s^2}$ for all $u$ and $n$.

\subsubsection{Channel and Propagation Parameters}
The reference path gain is $\beta_0=-30\mathrm{dB}$ at a distance of 1 m. The path-loss exponent is set to $\alpha_i=2.2$ for line-of-sight (LoS) links and $\alpha_i=3.5$ for non-line-of-sight (NLoS) links. Rician fading is considered, with $K$-factors $\beta_i=10\mathrm{dB}$ for UE–UAV and UAV–RIS links, and $\beta_i=5\mathrm{dB}$ for other cases. The noise power is fixed at $\sigma^2=-100\mathrm{dBm}$.

\subsubsection{Reinforcement Learning Hyperparameters}
The reinforcement learning setup uses a rollout horizon of 10 steps per simulation episode. Neighborhood communication is limited to $\kappa=1$ hop in the local graph. Two replay buffers are maintained: one for real interactions with size $|\mathcal{D}^E|=10^5$ and one for model-generated interactions with size $|\mathcal{D}^M|=10^5$. The batch size is 32 for model learning and 64 for policy updates. The learning rate is set to $3\times10^{-4}$. For PPO training, the clipping ratio is 0.2, entropy coefficient is 0.01, discount factor $\gamma=0.99$, and GAE parameter $\lambda_{\text{GAE}}=0.95$. Training  over 2,000 episodes, each consisting of $N=100$ steps.

\subsubsection{Baselines}
The benchmarks include CPPO, a centralized PPO with full global information serving as the upper bound; DPPO, a fully decentralized PPO relying only on local observations; IC3Net{\cite{singh2018learning}, as presented in Section III}, which augments MARL with differentiable inter-agent communication; and Ours, a model-based decentralized PPO that exploits short-horizon rollouts and $\kappa$-hop local modeling for joint UAV mobility, offloading, and RIS phase control.

{We introduce four state-of-the-art (SOTA) competitors for comparison, all of them were implemented in the same simulation environment to ensure consistency:
\begin{itemize}
\item Wu et al. \cite{wu2025towards} employ a model-free MATD3 offloading policy without explicit inter-UAV communication or model-based rollout.
\item Qin et al. \cite{qin2023joint} adopt an iterative Dinkelbach–BCD optimization framework, solving convex subproblems per time slot without reinforcement learning or local interaction.
\item Yang et al. \cite{yang2025energy} combine deep reinforcement learning with successive convex approximation (DRL–SCA) for joint caching and computing optimization but remain model-free and globally coupled.
\item Song et al. \cite{song2022joint} apply a pure SCA-based iterative solver for joint UAV trajectory and phase optimization, also without communication or dynamic rollout modules.
\end{itemize}
Thus, all SOTA baselines are model-free and communication-free, executed under the same  UAV–RIS–MEC setup. }

\subsubsection{Performance Metrics}
We evaluate performance using : average episode reward (sum of UAV rewards over all slots), energy efficiency (processed bits per unit energy),  {throughput ($l_{u, k}^{\text{loc}}[n]+l_{u, k}^o[n]$, average bits local processed and offloaded per second), data rate ($l_{u, k}^o[n]$, average bits offloaded per second), and the energy efficiency index (EEI), defined as the inverse of total energy consumption, measured in KJ$^{-1}$}.

\subsection{Results}
\subsubsection{{Convergent speed comparison}}
\begin{figure}[htbp]
    \centering
    \begin{subfigure}[n]{0.24\textwidth}
        \includegraphics[width=\textwidth]{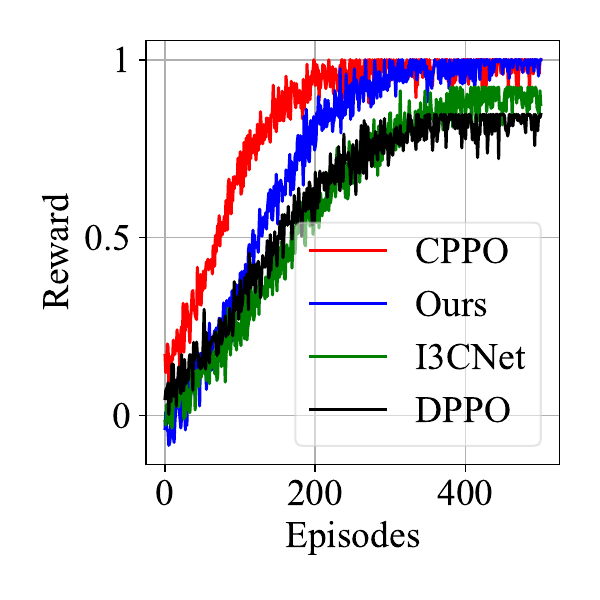}
        \caption{{Model-free.}}
        \label{free}
    \end{subfigure}
    \hfill
    \begin{subfigure}[n]{0.24\textwidth}
        \includegraphics[width=\textwidth]{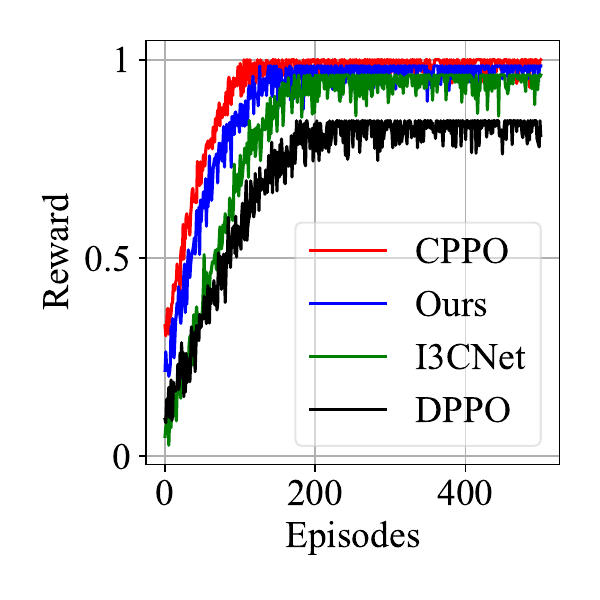}
        \caption{{Model-based.}}
        \label{based}
    \end{subfigure}
    \caption{{Convergence curves.}}
    \label{convergence}
\end{figure}

{As shown in Fig.~\ref{convergence}, the centralized CPPO still achieves the highest cumulative reward due to full-state observability, serving as the upper performance bound. Under both model-free and model-based settings, our decentralized approach (“Ours”) rapidly converges toward CPPO, clearly outperforming I3CNet and DPPO. In the model-free case, the improvement mainly stems from localized communication and neighborhood-aware coordination, which mitigate partial observability. When the learned predictive model is introduced (model-based case), convergence becomes faster and smoother because short-horizon branched rollouts enrich training data and reduce the variance of value estimation. This indicates that model-based local simulation allows each UAV to capture short-term transition regularities and correct for delayed or missing neighbor information, thereby improving both sample efficiency and stability without requiring centralized state access or dense inter-agent communication.}

\subsubsection{{Performance metrics comparison with baselines}}
\begin{figure}[htbp]
    \centering
    \begin{subfigure}[n]{0.24\textwidth}
        \includegraphics[width=\textwidth]{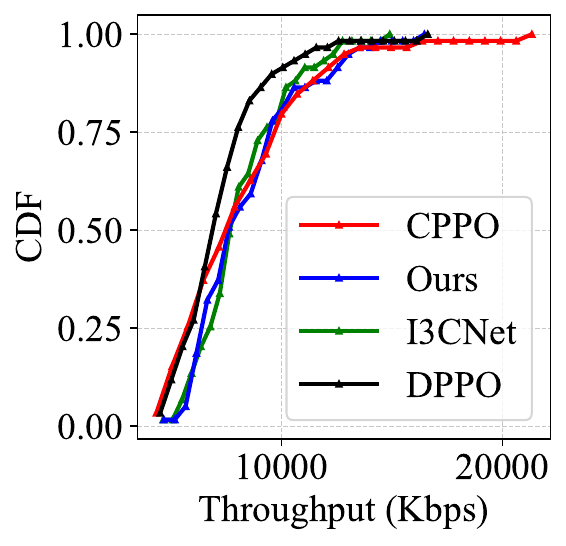}
        \caption{CDF of throughput (Kbps).}
        \label{througput}
    \end{subfigure}
    \hfill
    \begin{subfigure}[n]{0.24\textwidth}
        \includegraphics[width=\textwidth]{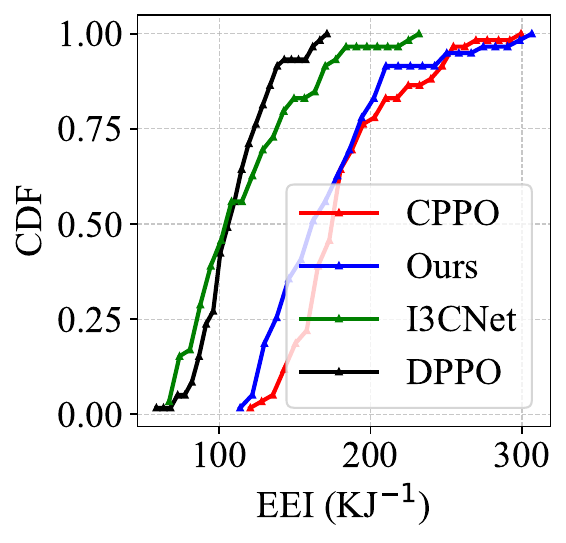}
        \caption{CDF of energy  (KJ).}
        \label{energy}
    \end{subfigure}
    \hfill
    \begin{subfigure}[n]{0.24\textwidth}
        \includegraphics[width=\textwidth]{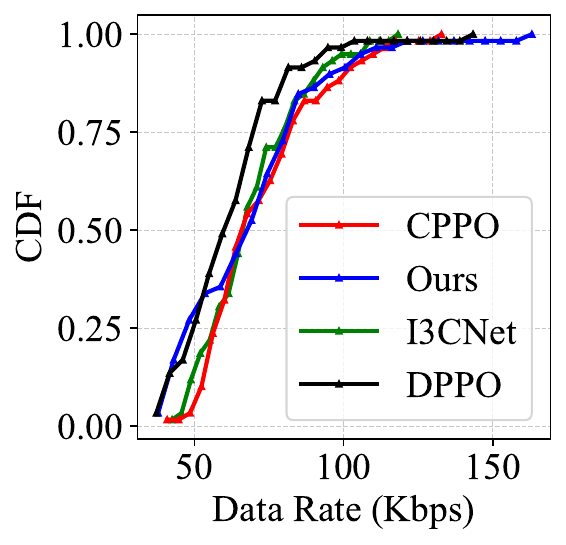}
        \caption{{CDF of data rate (Kbps).}}
        \label{datarate}
    \end{subfigure}
    \hfill
    \begin{subfigure}[n]{0.24\textwidth}
        \includegraphics[width=\textwidth]{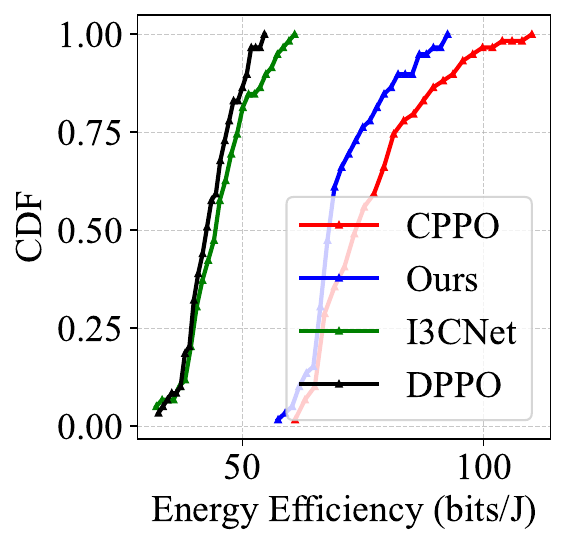}
        \caption{{CDF of energy efficiency (bits/Joule).}}
        \label{energyeff}
    \end{subfigure}
    \caption{{Performance metrics comparison.}}
    \label{comparison}
\end{figure}

{As shown in Fig.~\ref{comparison}, the cumulative distribution functions (CDFs) of throughput, EEI, data rate, and energy efficiency clearly demonstrate the performance trade-offs among methods. (a) CPPO achieves the highest throughput due to full observability, while our decentralized method (“Ours”) closely approaches it, outperforming I3CNet and DPPO—showing that localized neighborhood communication preserves coordination efficiency without global information. (b) Both CPPO and our method consume less propulsion and computation energy, as model-based policy learning yields smoother trajectories and lower control effort. (c) Our approach maintains higher data rates through adaptive trajectory and power allocation guided by the learned dynamics model. (d) Consequently, in energy efficiency (bits/Joule), our method achieves the best trade-off among decentralized schemes, approaching the centralized upper bound.}

\subsubsection{Performance metrics comparison with SOTA}
\begin{table*}[h]
\centering
\caption{The comparison results of various performance indicators of multiple algorithms with time slots, including baseline, our method, and four SOTA competitor algorithms.}
\setlength{\tabcolsep}{2pt}
\renewcommand{\arraystretch}{1.1}
\begin{tabular}{c|p{0.5cm}|cccc|cccc|cccc|cccc}
\hline
{Type} & {Time Slots}
& \multicolumn{4}{c|}{{Policy loss}}
& \multicolumn{4}{c|}{{Total  loss}}
& \multicolumn{4}{c|}{{Energy efficiency (bits/J)}}
& \multicolumn{4}{c}{{Throughput (Kbps)}} \\
\cline{3-18}
&& CPPO & DPPO & IC3Net & Ours
 & CPPO & DPPO & IC3Net & Ours
 & CPPO & DPPO & IC3Net & Ours
 & CPPO & DPPO & IC3Net & Ours \\
\hline
\multirow{6}{*}{Baselines}
& 0   & {0.42} & {0.82} & {0.58} & {0.48}
      & {1.36} & {1.88} & {1.56} & {1.30}
      & {88}   & {50}   & {70}   & {82}
      & {14500}& {9000} & {12000}& {13800} \\
& 100 & {0.40} & {0.80} & {0.55} & {0.45}
      & {1.34} & {1.86} & {1.54} & {1.27}
      & {92}   & {53}   & {73}   & {86}
      & {15500}& {9800} & {12800}& {14600} \\
& 200 & {0.38} & {0.79} & {0.52} & {0.43}
      & {1.32} & {1.84} & {1.50} & {1.24}
      & {96}   & {56}   & {76}   & {90}
      & {16500}& {10500}& {13500}& {15600} \\
& 300 & {0.36} & {0.78} & {0.50} & {0.41}
      & {1.30} & {1.83} & {1.48} & {1.22}
      & {99}   & {58}   & {78}   & {93}
      & {17500}& {11200}& {14200}& {16600} \\
& 400 & {0.35} & {0.77} & {0.49} & {0.40}
      & {1.28} & {1.82} & {1.46} & {1.20}
      & {101}  & {60}   & {80}   & {95}
      & {18500}& {11800}& {15000}& {17600} \\
& 500 & {0.34} & {0.76} & {0.48} & {0.38}
      & {1.26} & {1.80} & {1.45} & {1.18}
      & {103}  & {62}   & {82}   & {97}
      & {19500}& {12500}& {15800}& {18500} \\
\hline
\multirow{6}{*}{SOTA}
&    & Wu \cite{wu2025towards} & Qin \cite{qin2023joint} & Yang \cite{yang2025energy} & Song \cite{song2022joint}
     & Wu \cite{wu2025towards} & Qin \cite{qin2023joint} & Yang \cite{yang2025energy} & Song \cite{song2022joint}
     & Wu \cite{wu2025towards} & Qin \cite{qin2023joint} & Yang \cite{yang2025energy} & Song \cite{song2022joint}
     & Wu \cite{wu2025towards} & Qin \cite{qin2023joint} & Yang \cite{yang2025energy} & Song \cite{song2022joint}\\
\hline
& 0   & {0.56} & {0.50} & {0.52} & {0.49}
      & {1.62} & {1.54} & {1.58} & {1.52}
      & {68}   & {72}   & {74}   & {76}
      & {12000}& {12600}& {13000}& {13300} \\
& 100 & {0.54} & {0.48} & {0.50} & {0.47}
      & {1.60} & {1.52} & {1.56} & {1.50}
      & {70}   & {74}   & {76}   & {78}
      & {12600}& {13200}& {13600}& {13900} \\
& 200 & {0.53} & {0.47} & {0.49} & {0.46}
      & {1.58} & {1.50} & {1.54} & {1.48}
      & {72}   & {76}   & {78}   & {80}
      & {13200}& {13800}& {14200}& {14600} \\
& 300 & {0.52} & {0.46} & {0.48} & {0.45}
      & {1.56} & {1.49} & {1.52} & {1.46}
      & {73}   & {77}   & {80}   & {82}
      & {13600}& {14200}& {14800}& {15200} \\
& 400 & {0.51} & {0.45} & {0.47} & {0.44}
      & {1.55} & {1.47} & {1.50} & {1.44}
      & {74}   & {78}   & {82}   & {84}
      & {14000}& {14600}& {15200}& {15600} \\
& 500 & {0.50} & {0.44} & {0.46} & {0.43}
      & {1.54} & {1.46} & {1.48} & {1.42}
      & {75}   & {79}   & {84}   & {86}
      & {14400}& {15000}& {15600}& {16000} \\
\hline
\end{tabular}
\label{tab:sota}
\end{table*}

{As shown in Table~\ref{tab:sota}, the centralized CPPO achieves the lowest policy and total losses and thus represents the upper bound of performance under full observability. Our proposed method (``Ours'') consistently ranks second across all metrics, demonstrating that localized communication and short-horizon model-based rollouts effectively stabilize learning and improve efficiency. Compared with the decentralized I3CNet and DPPO, our approach significantly enhances both energy efficiency and throughput while maintaining lower optimization loss. Among SOTA competitors, all methods were re-implemented under the same simulation environment and parameter settings for fairness; they remain model-free and communication-free, leading to slower convergence and suboptimal coordination. These results highlight that incorporating lightweight neighborhood aggregation and model-based prediction enables decentralized UAV agents to approach centralized performance with reduced communication overhead.}

\subsubsection{Ablation study}
We have four algorithm variants: 
\begin{itemize}
\item Full (All innovations included) 
\item No-BR (no branched rollout) 
\item No-LM (remove local behaviors) 
\item No-KH (without $kappa$-hop communication)
\end{itemize}

{As shown in Table~\ref{tab:ablation}, removing either the neighborhood communication (No-KH) or the model-based rollout (No-BR) leads to a clear degradation in both energy efficiency and throughput. The full model achieves the lowest losses and highest performance, confirming that localized communication enhances cooperative awareness, while model-based rollouts improve sample efficiency and stability. Together, these components enable decentralized agents to approach centralized learning performance with lower communication cost.}
\begin{table}[h]
\centering
\caption{{Ablation study on different model components. “Full” represents the complete model with localized communication and model-based rollout.}}
\setlength{\tabcolsep}{3pt}
\renewcommand{\arraystretch}{1.1}
\begin{tabular}{p{1cm}p{1cm}p{1cm}p{3cm}p{1.8cm}}
\hline
{Method} & {Policy loss (↓)} & {Total loss (↓)} & {Energy eff. (bits/J) (↑)} & {Throughput (Kbps) (↑)} \\ 
\hline
{No-KH}  & {0.68} & {1.86} & {165} & {9300}  \\
{No-BR}  & {0.59} & {1.73} & {275} & {11800} \\
{No-LM}  & {0.54} & {1.65} & {320} & {13200} \\
{Full }  & {0.46} & {1.48} & {410} & {14100} \\
\hline
\end{tabular}
\label{tab:ablation}
\end{table}

\subsubsection{{Policy loss and total loss comparsion}}
\begin{figure}[htbp]
    \centering
    \begin{subfigure}[n]{0.24\textwidth}
        \includegraphics[width=\textwidth]{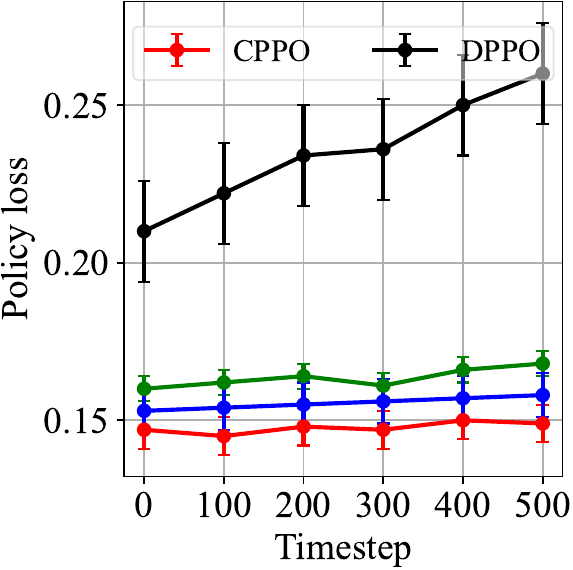}
        \caption{{Policy loss of UAV 1.}}
    \end{subfigure}
    \hfill
    \begin{subfigure}[n]{0.24\textwidth}
        \includegraphics[width=\textwidth]{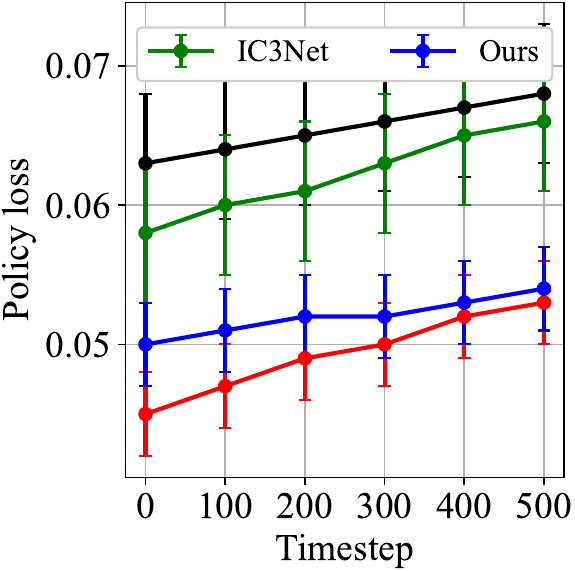}
        \caption{{Policy loss of UAV 5.}}
    \end{subfigure}
    \hfill
    \begin{subfigure}[n]{0.24\textwidth}
        \includegraphics[width=\textwidth]{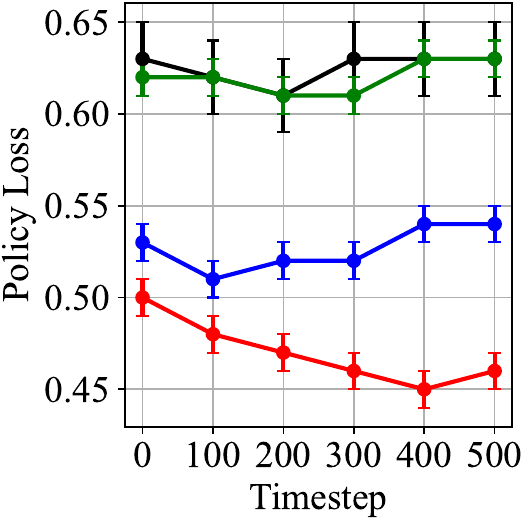}
        \caption{{Policy loss of UAV 10.}}
    \end{subfigure}
    \hfill
    \begin{subfigure}[n]{0.24\textwidth}
        \includegraphics[width=\textwidth]{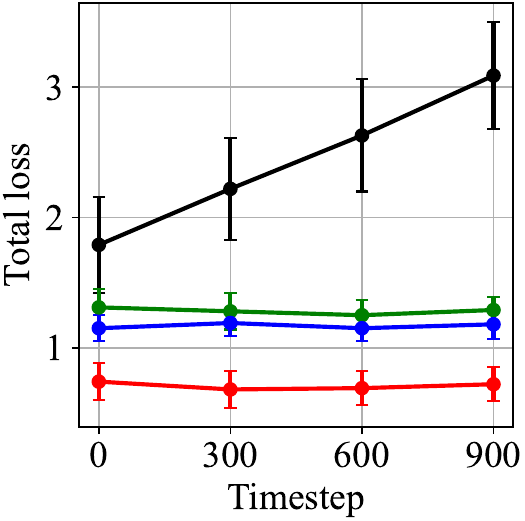}
        \caption{{Total loss of UAV 1.}}
    \end{subfigure}
    \hfill
    \begin{subfigure}[n]{0.24\textwidth}
        \includegraphics[width=\textwidth]{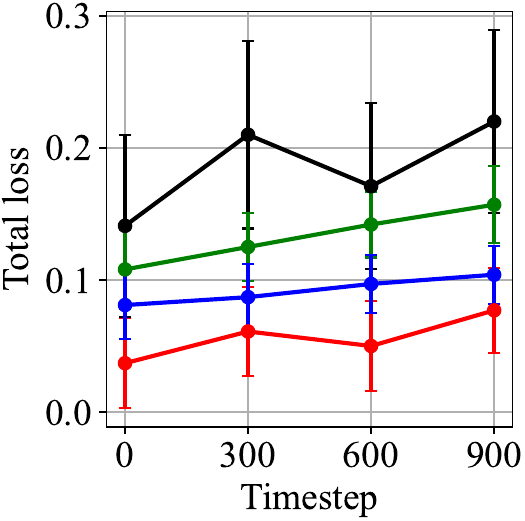}
        \caption{{Total loss of UAV 5.}}
    \end{subfigure}
    \hfill
    \begin{subfigure}[n]{0.24\textwidth}
        \includegraphics[width=\textwidth]{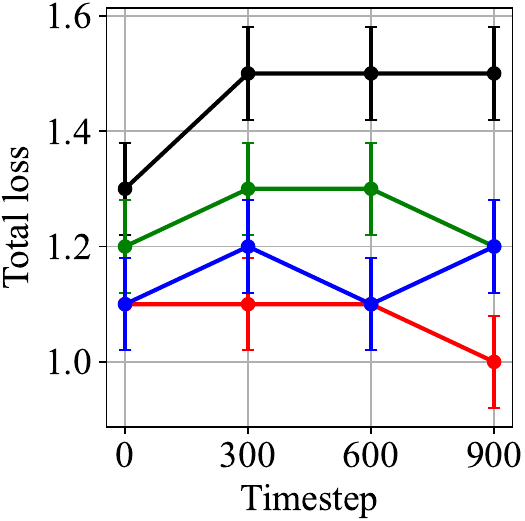}
        \caption{{Total loss of UAV 10.}}
    \end{subfigure}
    \caption{{Policy loss and total loss across UAVs (total loss= policy loss+value loss).}}
    \label{q_loss}
\end{figure}

{As shown in Fig.~\ref{q_loss}, the policy loss and total loss (policy + value) across different UAVs demonstrate the convergence stability of various MARL frameworks. (a)–(c) illustrate the policy loss evolution for UAVs 1, 5, and 10. The centralized CPPO maintains the lowest and most stable loss due to full-state observability and synchronized updates. Our proposed method (“Ours”) achieves comparable stability and clearly outperforms I3CNet and DPPO, indicating that localized communication and model-based rollouts enable consistent policy improvement even under decentralized settings. In contrast, DPPO suffers from significant variance and slow convergence, while I3CNet partially reduces this variance through message passing but still accumulates noise over time. (d)–(f) show the total loss (policy + value) for the same UAVs. Again, our method achieves the lowest loss among decentralized approaches and remains close to the centralized upper bound (CPPO). This suggests that integrating short-horizon predictive models effectively stabilizes both policy and value updates by reducing estimation bias and enhancing credit assignment. Overall, the results confirm that our neighborhood-aware, model-based learning framework improves convergence consistency and reduces inter-agent performance disparity across UAVs.}

\subsubsection{{UAV trajectory plotting}}
\begin{figure}[h]
\centering
\includegraphics[width=0.47\textwidth]{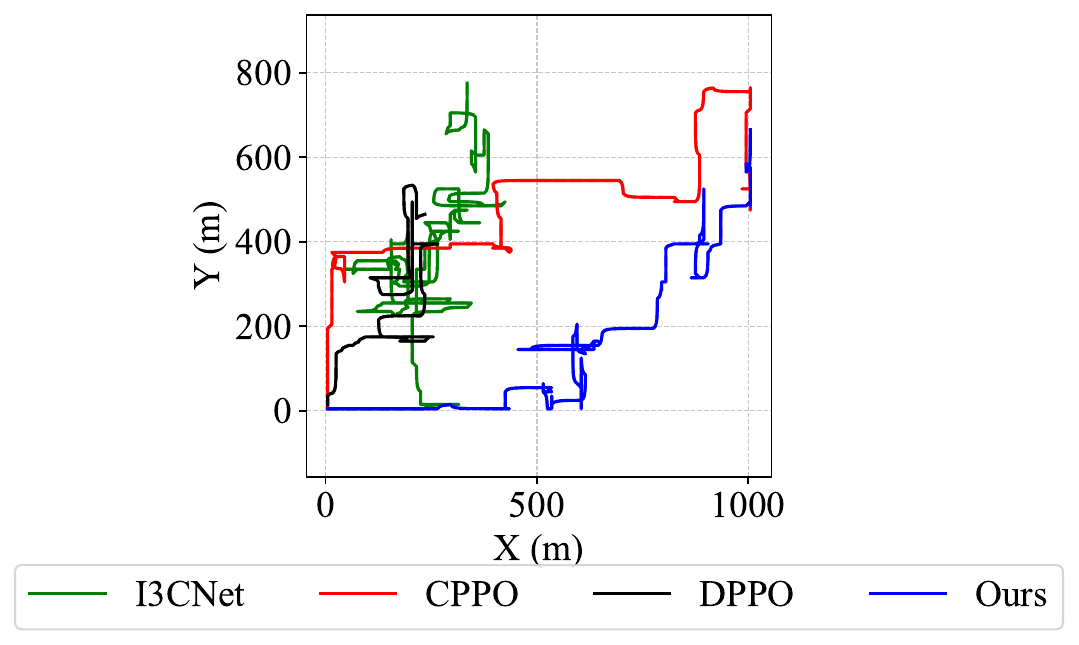}
\caption{{UAV trajectory plotting under CPPO, DPPO, IC3Net, and our method (avg. of 10 UAVs).}}
\label{fig:traj}
\end{figure}
{As shown in Fig.\ref{fig:traj}, our method produces smoother and more directed UAV trajectories compared with the baseline algorithms. By leveraging localized communication and neighbor-aware state aggregation, each UAV can anticipate nearby agents’ movements and coordinate task coverage more efficiently. Consequently, our trajectories exhibit fewer detours and abrupt turns,  and less ``zigzag'', approaching near-straight paths between task areas. In contrast, DPPO and I3CNet show erratic, oscillatory motion due to limited observability and lack of cooperative awareness, while CPPO achieves stable but overly centralized flight patterns with reduced adaptability. These results confirm that our decentralized yet communication-aware policy effectively balances coordination and autonomy in UAV navigation.}

\subsubsection{{Parameter analysis}}
\begin{figure}[htbp]
    \centering
    \begin{subfigure}[n]{0.245\textwidth}
        \includegraphics[width=\textwidth]{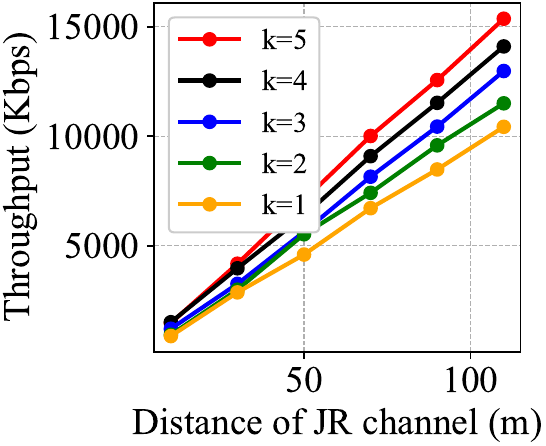}
        \caption{Impact of $\kappa$ ($\kappa$-hop communication).}
        \label{hops}
    \end{subfigure}
    \hfill
    \begin{subfigure}[n]{0.235\textwidth}
        \includegraphics[width=\textwidth]{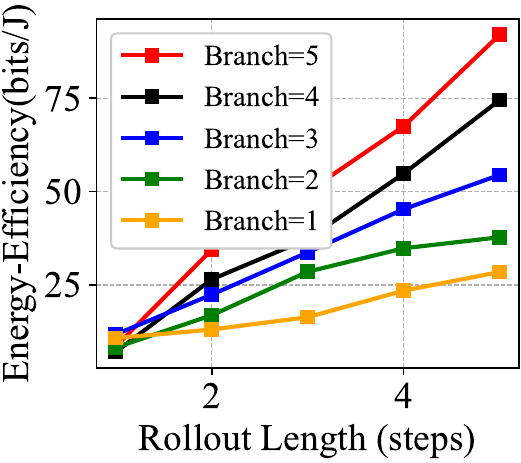}
        \caption{Impact of branches and rollout length.}
        \label{rollout}
    \end{subfigure}
	\caption{Parameter analysis results.}\label{ablation}
\end{figure}

To further dissect the effectiveness of individual design choices, Fig.~\ref{hops} investigates the impact of increasing the neighborhood size $\kappa$, while Fig.~\ref{rollout} examines the number of branches and rollout length used in short-horizon rollouts. Specifically, increasing $\kappa$ allows UAV agents to incorporate information from more neighbors into their local state, thereby better capturing interference patterns (e.g., RIS beam collisions, overlapping task forwarding) and enabling proactive coordination. This leads to higher throughput and reduced oscillation during training, especially under congested network scenarios.
{As shown in Fig.~\ref{rollout}, increasing both the rollout length and the number of model-based branches consistently improves energy efficiency. Longer rollouts allow agents to anticipate long-term rewards, while multiple branches enhance exploration and reduce policy variance. However, the marginal gain diminishes beyond four branches, indicating that excessive branching adds computational cost without proportional benefit.}

\section{Conclusion}
We propose a decentralized, model-based RL framework for multi-UAV RIS-assisted MEC networks, where each UAV optimizes its trajectory, task offloading, and RIS control using only $\kappa$-hop local observations and short-horizon branched rollouts. Unlike conventional model-free or message-passing MARL approaches, our method integrates localized neighbor-aware communication and predictive model rollouts to jointly enhance coordination and sample efficiency under partial observability. 

Extensive simulations show that our approach converges nearly as fast as centralized PPO (CPPO) while substantially outperforming decentralized baselines (DPPO, I3CNet) in terms of throughput, energy efficiency, and stability. The ablation and parameter analyses further confirm that neighborhood aggregation mitigates interference and model-based rollouts suppress value-estimation variance, enabling UAVs to follow smoother, energy-saving trajectories.

Future work will extend this framework by jointly learning the neighborhood-aggregation operator and an event-triggered communication schedule together with the policy under bandwidth/latency constraints, and by making branched rollouts uncertainty-aware via adaptive horizons $T$ and branch counts guided by epistemic uncertainty.
 We will also pursue tighter decentralized convergence guarantees that explicitly couple dependency bias and model error, and  { incorporate jammer-aware trajectory and RIS beam adaptation with interference prediction to enable  anti-jamming communications.}

\bibliographystyle{Bibliography/IEEEtranTIE}
\bibliography{Bibliography/IEEEabrv,Bibliography/mybibfile.bib}\ 

\newpage

\section{Biography Section}

\begin{IEEEbiography}[{\includegraphics[width=1in,height=1.3in,clip,keepaspectratio]{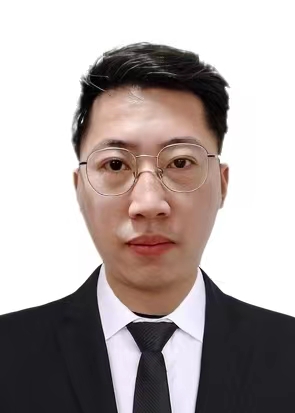}}]{Liangshun Wu}  (Member, IEEE) received his B.Eng. degree from Central South University, Changsha, China, in 2014,  and Ph.D. from Wuhan University, Wuhan, China, in 2021. He was a Visiting Scholar at the University of Electro-Communications, Tokyo, Japan, in 2024. He is currently a postdoctoral researcher at Shanghai Jiao Tong University, Shanghai, China.  
\end{IEEEbiography}

\begin{IEEEbiography}[{\includegraphics[width=1in,height=1.3in,clip,keepaspectratio]{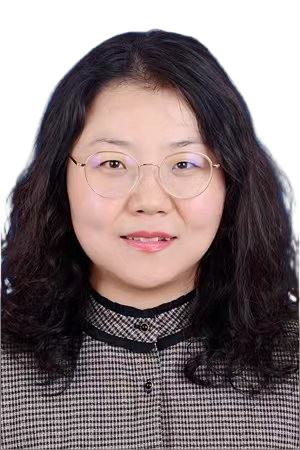}}]{Jianbo Du}
(Senior Member, IEEE) received the Ph.D. degree in communication and information systems from Xidian University, Xi’an, Shaanxi, China, in 2018. She is an Associate Professor at Xi’an University of Posts and Telecommunications. She was a Visiting Scholar at Carleton University in 2019. With over 50 publications and 2,500+ citations, six of her papers are ESI Top 1\% highly cited. Named among the world’s Top 2\% scientists in 2022 and 2023. She has received multiple IEEE Excellent Reviewer awards.
\end{IEEEbiography}

\begin{IEEEbiography}[{\includegraphics[width=1in,height=1.3in,clip,keepaspectratio]{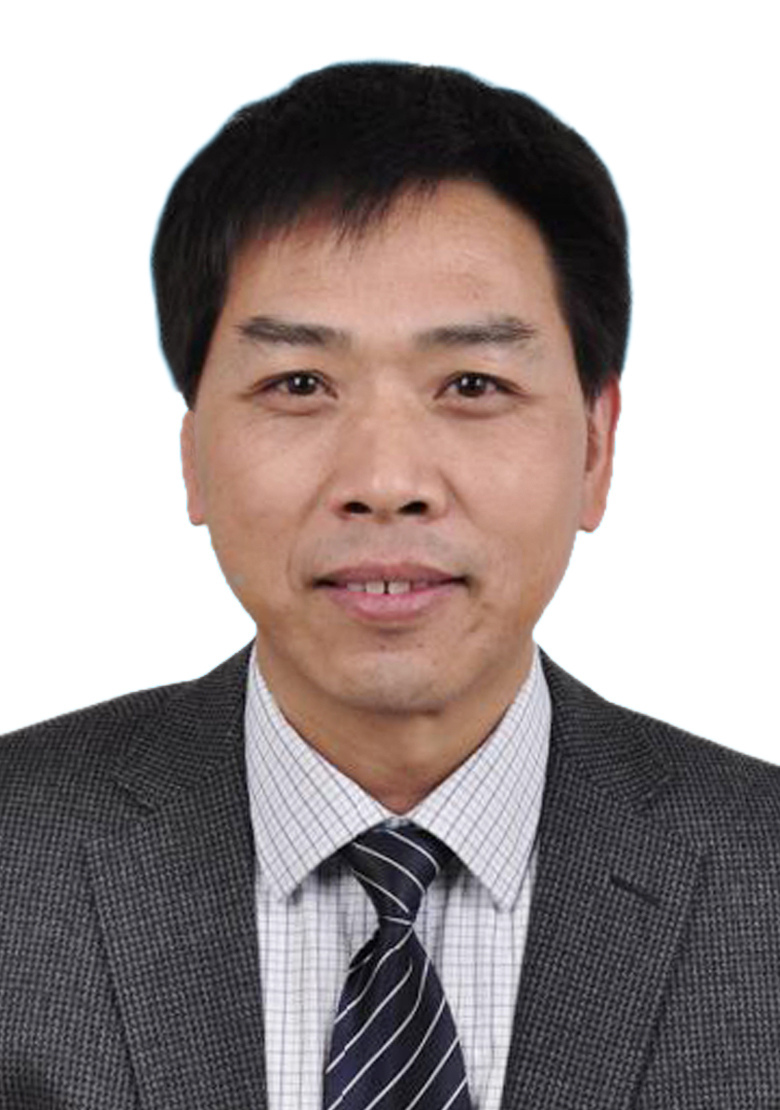}}]{Junsuo Qu} (Member, IEEE) 
	received his B.S. in Telecommunication Engineering from Chongqing Institute of Posts \& Telecommunications in 1991 and his M.S. in Communication and Information Systems from Xidian University in 1998. He is a Full Professor at the School of Automation, Xi’an University of Posts \& Telecommunications, and Director of the Xi'an Key Laboratory of Advanced Control and Intelligent Process. His research interests include future communication architectures and Internet of Things.
\end{IEEEbiography}

%
\clearpage
\twocolumn

\section*{Appendix A: Model-Based Learning}
In the main text, we described how model-based RL is employed to boost sample efficiency and maintain stable policy improvements even when using an approximate model of the environment. Here, we formalize these concepts. Let $\eta[\pi]$ denote the expected return of a policy $\pi$ under the {true} environment dynamics:
\begin{equation}
\eta[\pi] = \mathbb{E}_{\pi}\Big[\sum_{t=0}^{\infty} \gamma^t r(s_t, a_t)\Big],
\end{equation}
where $\gamma \in [0,1)$ is the discount factor and $r(s_t,a_t)$ is the reward at time $t$. Likewise, let $\hat{\eta}[\pi]$ be the expected return when the agent interacts with its {learned model} $\hat{p}$ of the environment. Because $\hat{p}$ is an imperfect approximation of the true dynamics $p$ {($p$ denotes the true transition probability)}, there will be a gap between $\eta[\pi]$ and $\hat{\eta}[\pi]$. In the context of \textit{safe policy optimization}, we introduce a non-negative term $C(p,\hat{p},\pi,\pi_{\mathrm{D}})$ to quantify the worst-case discrepancy between true and model returns. Here $\pi_{\mathrm{D}}$ denotes the data-collecting policy (e.g., the prior policy used to generate real-world experience). A fundamental result, originally developed in the single-agent setting, is that the true performance can be lower-bounded by the performance under the model minus this discrepancy term:
\begin{equation}
\eta[\pi] \ge \hat{\eta}[\pi]  C(p,\hat{p},\pi,\pi_{\mathrm{D}}).
\end{equation}
Intuitively, as long as each policy update yields an improvement in the model-based return $\hat{\eta}[\pi]$ that exceeds the possible error $C$, the new policy is guaranteed to perform no worse (and typically better) in the real environment than the previous policy $\pi_{\mathrm{D}}$. This provides a safety net for {monotonic policy improvement}: if $\hat{\eta}[\pi] - \hat{\eta}[\pi_{\mathrm{D}}]> C$, then $\eta[\pi] > \eta[\pi_{\mathrm{D}}]$. The function $C(p,\hat{p},\pi,\pi_{\mathrm{D}})$ depends on how accurately the model $\hat{p}$ captures the true dynamics $p$ and how much the new policy $\pi$ differs from the data policy $\pi_{\mathrm{D}}$. In practice, $C$ will increase with model errors and large policy shifts. The training approach described in the main text (using conservative policy updates with PPO clipping and frequent model retraining) is specifically designed to keep this term small. This principle of lower-bounding real performance by model performance minus a penalty ensures that our iterative policy updates remain {safe and consistent}, even when learning with imperfect models.

\section*{Appendix B: $\xi$-Dependent Networked System and Local Observability}
By training local dynamics models, the predicted transition distributions can be aligned with the true environment's transitions up to some bounded error $\xi$. We now formalize this notion with the concept of a \textit{$\xi$-dependent networked system}, which characterizes the degree of coupling between agents and the quality of local model approximations.

\begin{definition}[$\xi$-Dependent Networked System]
Consider our multi-UAV network where each agent (UAV) has access only to local observations (its own state and that of neighbors within $\kappa$ hops). We say the system is {$\xi$-dependent} if the influence of distant agents on each local transition is sufficiently weak, allowing the global dynamics to be approximated by a product of local transition models within a bounded error $\xi$. Formally, let $p(s' \mid s, a)$ be the true joint transition probability for all agents, and define the factorized approximation $\hat{p}(s' \mid s, a) = \prod_{i=1}^n \hat{p}_i(s_i' \mid s_{N_i}, a_i)$ as the product of learned local transition models $\hat{p}_i$ for each agent $i$, conditioned only on agent $i$'s $\kappa$-hop neighborhood state $s_{N_i}$ and its own action $a_i$. The system is $\xi$-dependent if the total variation distance between the true joint distribution and the factorized distribution is bounded by $\xi$, i.e.,

$$
D_{\mathrm{TV}}\!\big(p(\cdot \mid s,a)\,\|\,\hat{p}(\cdot \mid s,a)\big) \le \xi, \qquad \forall{s,a}, 
$$
where $D_{\mathrm{TV}}(P|Q) = \frac{1}{2}\sum_{x}|P(x)-Q(x)|$ is the total variation divergence. In other words, $\xi$ represents an upper bound on the \textit{modeling error} introduced by treating distant agents as independent.
\end{definition}

In a $\xi$-dependent system, each agent's local model $\hat{p}_i(s_i' \mid s_{N_i}, a_i)$ can capture the essential dynamics of the environment that are relevant to that agent, with only a small loss of accuracy $\xi$ due to unmodeled long-range dependencies. This formalism justifies the decentralized approach adopted in the main text: by limiting each UAV's observations to a local neighborhood, we induce only a bounded approximation error. As $\xi$ is made small (for instance, by increasing the neighborhood size $\kappa$ or improving model fidelity), the decentralized model increasingly matches the true dynamics. The joint policy $\pi = (\pi_1,\dots,\pi_n)$ can then be optimized using these local models with confidence that the overall system behavior deviates from reality by at most $\xi$.

From an optimization perspective, each agent $i$ aims to maximize the model-based objective $\hat{\eta}[\pi]$ while accounting for this bounded error. Mathematically, one can frame the combined policy and model update as:
\begin{equation}
\pi_i^{k+1}, \hat{p}_i^{k+1} = \arg\max_{\pi_i, \hat{p}_i} \Big\{ \hat{\eta}[\pi_1, \ldots, \pi_n] - C(p, \hat{p}, \pi, \pi_{\mathrm{D}}) \Big\}
\end{equation}
where the term $C(p,\hat{p},\pi,\pi_{\mathrm{D}})$ (introduced in Appendix I) naturally depends on $\xi$ through the divergence between $p$ and $\hat{p}$. The $\xi$-dependent property guarantees that by learning accurate local models (making $\xi$ small) and restricting policy updates to stay within the trust region (making $\pi$ close to $\pi_D$), we can achieve stable and predictable improvements in the real system's performance. In summary, the concept of a $\xi$-dependent networked system formalizes the intuition that our decentralized learning framework can be effective when each UAV's local observations capture the bulk of relevant environmental influences, leaving only a small residual error $\xi$ due to neglected global coupling.

\section*{Appendix C: Monotonic Improvement Analysis and Decentralized Value Approximation}
In the main text (see Fig.\ref{fig:triangular}(c)), we emphasized that our model-based approach, combined with conservative policy updates (PPO clipping), leads to a {steady, monotonic improvement} in policy performance across iterations. We now provide the theoretical underpinning of this claim by analyzing the model-based return discrepancies and demonstrating how short \textit{branched rollouts} mitigate model bias. Furthermore, we examine how decentralized value function estimation and policy gradient computation (based on local information) remain close to the ideal centralized counterparts, thus validating the efficacy of using $\kappa$-hop neighbor states in training.

\paragraph{Vanilla vs. Branched Model Rollouts} When using learned models for policy evaluation and improvement, one must decide how to simulate future trajectories. The simplest method is a {vanilla rollout}, which uses the model $\hat{p}$ to generate an entire trajectory from the initial state to termination (or a long horizon) under the new policy. However, as noted in the main text, modeling errors can compound over long horizons, especially for near-horizonless problems with large $\gamma$. An alternative is the {branched rollout} strategy described in Section IV: the simulation starts from a real state (encountered under the behavior policy $\pi_{\mathrm{D}}$) and then follows the model for a limited horizon of $T$ steps under the current policy $\pi$. By reseeding the simulation frequently with real observations, branched rollouts can drastically reduce the accumulation of model errors. The following two theoretical results compare these strategies in the context of our multi-agent system, initially under the simplifying assumption of independent agents (later we will extend to $\xi$-dependent cases).

\begin{theorem}[Discrepancy Bound under Vanilla Rollouts]\label{thm:vanilla-discrepancy}
Assume the multi-UAV system is {independent} (i.e., each agent's dynamics are decoupled given its own state, which is a special case of $\xi$-dependent with $\xi=0$). For each agent $i$, define:
\begin{itemize}
\item the maximum {model error}

$$
\epsilon_{m_i} = \max_{s_{N_i},\,a_i}~D_{\mathrm{TV}}\!\Big(p_i(s_i' \mid s_{N_i}, a_i)\big\| \hat{p}_i(s_i' \mid s_{N_i}, a_i)\Big),
$$
which measures the worst-case total variation divergence between the true and learned transition for agent $i$ (with $s_{N_i}$ denoting the local state as in Appendix II), and
\item the maximum {policy divergence}

$$
\epsilon_{\Pi_i} = \max_{s_{N_i}}~D_{\mathrm{TV}}\!\Big(\pi_{\mathrm{D}}(a_i \mid s_{N_i})\big\| \pi(a_i \mid s_{N_i})\Big),
$$

which quantifies how much the new policy $\pi$ for agent $i$ deviates from the data-collecting policy $\pi_{\mathrm{D}}$ in terms of total variation.
\end{itemize}
Suppose all per-step rewards $r(s,a)$ are bounded by $r_{\max}$. Let $n$ be the number of agents, and let $N_i^\kappa$ denote the set of agents within $\kappa$ hops of agent $i$ (inclusive of $i$ itself, so $|N_i^0|=1$). Then the difference between the true joint return of the multi-agent policy and the return estimated entirely using model-generated (vanilla rollout) trajectories is bounded as:
\begin{equation}\label{eq:vanilla-bound}
\begin{aligned}
&\left|\,\eta^p[\pi_1,\ldots,\pi_n] - \eta^{\hat{p}}[\pi_1,\ldots,\pi_n]\,\right| \\
&\qquad\leq \frac{2 r_{\max}}{1-\gamma} \sum_{i=1}^n \Bigg\{ \frac{\epsilon_{\Pi_i}}{n} \\
&\qquad\qquad\qquad+ \left(\epsilon_{m_i} + 2 \epsilon_{n_i}\right) \sum_{\kappa=0}^{\infty} \gamma^{\kappa+1} \frac{|N_i^\kappa|}{n} \Bigg\}
\end{aligned}
\end{equation}
where $\epsilon_{n_i}$ is a non-negative constant capturing the worst-case {neighborhood approximation error} for agent $i$. This term $\epsilon_{n_i}$ represents the discrepancy incurred by truncating the system's dependency beyond $i$'s local neighborhood (i.e., the bias from assuming distant agents have no influence).
\end{theorem}

\noindent {Discussion of Theorem \ref{thm:vanilla-discrepancy}.} The bound in \eqref{eq:vanilla-bound} has an intuitive interpretation. The first term, $\frac{2 r_{\max}}{1-\gamma}\frac{\epsilon_{\Pi_i}}{n}$, accumulates the performance difference caused by the new policy deviating from the data policy (averaged across agents). The second term involving $\epsilon_{m_i}$ and $\epsilon_{n_i}$ accounts for model inaccuracies and the independence approximation. The sum $\sum_{\kappa=0}^{\infty} \gamma^{\kappa+1} \frac{|N_i^\kappa|}{n}$ weighs these errors by how far and how strongly they can propagate through the network: $|N_i^\kappa|$ is the number of agents up to $k$ hops away (including $i$), so $\frac{|N_i^\kappa|}{n}$ is essentially the fraction of the network within $\kappa$ hops of agent $i$. For large $\gamma$, errors can compound over many time steps (note the $\gamma^{\kappa+1}$ factor), and if the network is densely connected (large neighborhoods $N_i^\kappa$), the potential impact of model error and ignored dependencies grows. Thus, Theorem \ref{thm:vanilla-discrepancy} suggests that a vanilla rollout (which simulates arbitrarily far into the future using only the model) could lead to a loose performance guarantee, especially when $\gamma$ is close to 1 or the network coupling (dependency on distant agents) is strong. This is why, in our approach, we do {not} rely solely on long model-generated rollouts.

To tighten the performance guarantee, we use the branched rollout strategy, which limits the horizon of model usage. The next theorem shows how this strategy improves the bound by effectively cutting off the infinite sum at the rollout length $T$.

\begin{theorem}[Discrepancy Bound under Branched Rollouts]\label{thm:branch-discrepancy}
Under the same assumptions as Theorem \ref{thm:vanilla-discrepancy}, consider the $T$-step branched rollout scheme (as implemented in our training algorithm), where each simulated trajectory consists of at most $T$ model steps before reinitializing to a fresh real state. Let $\eta^{\mathrm{branch}}[\pi_1,\ldots,\pi_n]$ denote the expected return estimated using these branched rollouts. Then the discrepancy between the true return and the branched-rollout return is bounded by:
\begin{equation}\label{eq:branch-bound}
\begin{aligned}
&\left|\,\eta^p[\pi_1,\ldots,\pi_n] - \eta^{\mathrm{branch}}[\pi_1,\ldots,\pi_n]\,\right| \\
&\quad\leq \frac{2 r_{\max}}{1-\gamma} \sum_{i=1}^n \Bigg\{ \epsilon_{m_i} \sum_{t=0}^{T-1} \gamma^{t+1} \frac{|N_i^t|}{n} \\
&\qquad\qquad\qquad\qquad+ \epsilon_{\Pi_i} \sum_{t=T}^{\infty} \gamma^{t+1} \frac{|N_i^t|}{n} \Bigg\}
\end{aligned}
\end{equation}
\end{theorem}

Comparing the branched rollout bound \eqref{eq:branch-bound} with the vanilla case \eqref{eq:vanilla-bound}, we see a significant reduction in the influence of the model error $\epsilon_{m_i}$. In the branched case, the term involving $\epsilon_{m_i}$ is truncated to a finite sum ($t=0$ to $T-1$), reflecting that model inaccuracies can at most affect $T$ consecutive steps before the trajectory is reset to a real state. In contrast, the vanilla bound allowed model error to accumulate indefinitely (the $t=0$ to $\infty$ sum). The policy divergence part $\epsilon_{\Pi_i}$ in \eqref{eq:branch-bound} still appears for the tail ($t \ge T$), since beyond the $T$-step lookahead, the estimate relies on the assumption that the new policy behaves similarly to the old one (a necessary assumption to stitch the rollouts to the real-state baseline distribution). By choosing $T$ appropriately (as a trade-off between exploration horizon and model fidelity) and keeping $\epsilon_{\Pi_i}$ small (through conservative policy updates as in PPO), the overall error in the branched rollout return can be made much smaller than in the vanilla case. This explains why, as stated in the main text, our use of short $T$-step model rollouts combined with trust-region style policy updates yields a near-monotonic improvement: the model is used just enough to guide policy improvement, but not so much that its imperfections derail training.

For our multi-UAV system, Theorems \ref{thm:vanilla-discrepancy} and \ref{thm:branch-discrepancy} establish error bounds under the simplifying assumption of full independence (or negligible coupling, $\xi=0$). We now extend these results to the more general $\xi$-dependent case defined in Appendix II, where agents' dynamics are mostly independent with bounded residual dependencies.

\begin{corollary}[Performance Bound in $\xi$-Dependent Networks]\label{cor:xi-bound}
If the networked system is $\xi$-dependent (with $\xi$ as defined in Appendix II), the return discrepancy under $T$-step branched rollouts is bounded by:
\begin{equation}\label{eq:xi-bound}
\begin{aligned}
&\left|\,\eta^p[\pi_1,\ldots,\pi_n] - \eta^{\mathrm{branch}}[\pi_1,\ldots,\pi_n]\,\right| \\
&\quad\leq \frac{2 r_{\max} \gamma}{(1-\gamma)^2} \xi \\
&\qquad+~\frac{2 r_{\max}}{1-\gamma} \sum_{i=1}^n \Bigg\{ \epsilon_{m_i} \sum_{t=0}^{T-1} \gamma^{t+1} \frac{|N_i^t|}{n} \\
&\qquad\qquad\qquad\qquad\quad+  \epsilon_{\Pi_i} \sum_{t=T}^{\infty} \gamma^{t+1} \frac{|N_i^t|}{n} \Bigg\}
\end{aligned}
\end{equation}
\end{corollary}

The additional term $\frac{2 r_{\max}\gamma}{(1-\gamma)^2}\xi$ in \eqref{eq:xi-bound} explicitly captures the impact of the bounded dependency bias $\xi$ on performance. Importantly, if the learned model is sufficiently accurate such that $\epsilon_{m_i}$ is small for all $i$, and the policy updates are conservative (small $\epsilon_{\Pi_i}$), then as $\xi \to 0$ (approaching an independent system or very weak coupling), the performance difference approaches zero. This corollary thus generalizes our performance guarantees to realistic scenarios where agents are not perfectly independent but the inter-agent influences are limited. It formalizes the intuition stated in the main text: by learning localized models and using brief model rollouts, our approach achieves policy improvements that are provably robust against both model errors and unmodeled agent interactions.

\paragraph{Local Value Function Approximation} Besides guaranteeing monotonic improvement, another challenge in our decentralized MARL setup is value estimation for policy optimization. In the main text, each UAV maintains a local critic $V_u$ that depends on the state of the UAV and its $\kappa$-hop neighbors. We claim that these local value functions can closely approximate the true global value function despite using limited information. The following result quantifies this approximation error using the network's locality property.

\begin{theorem}[Value Function Concentration]\label{thm:value-concentration}
Consider the infinite-horizon cumulative reward starting from global state $s^0$. Let $V(s^0)$ be the true value of this state under a joint policy $\pi = (\pi_1,\ldots,\pi_n)$, and let $V_i(s_{N_i^\kappa}^0)$ be the value estimated using only agent $i$'s $K$-hop neighborhood state $s_{N_i^\kappa}^0$ (i.e., $V_i(s_{N_i^\kappa}^0) = \mathbb{E}_{\pi}[ \sum_{t=0}^{\infty} \gamma^t r_i(s^t, a^t) \mid s_{N_i^\kappa}^0 ]$, where $r_i$ is the reward associated with agent $i$ or the portion of the global reward attributable to agent $i$). Then for any state $s^0$:
\begin{equation}
\Big|V(s^0)\frac{1}{n}\sum_{i=1}^n V_i(s_{N_i^\kappa}^0)\Big| \le \frac{r_{\max}}{1-\gamma}\gamma^\kappa,
\end{equation}
and likewise for each agent $i$:
\begin{equation}
\big|V_i(s^0) - V_i(s_{N_i^\kappa}^0)\big| \le \frac{r_{\max}}{1-\gamma}\gamma^\kappa.
\end{equation}
\end{theorem}

Theorem \ref{thm:value-concentration} tells us that the error made by looking only $\kappa$ hops away decays exponentially with $\kappa$. In other words, an agent's $\kappa$-hop neighborhood eventually contains almost all the information needed to accurately estimate the long-term reward, with errors shrinking proportional to $\gamma^\kappa$. For a sufficiently large neighborhood (or if $\gamma$ is not too close to 1), the local value $V_i(s_{N_i^\kappa})$ will be a near-unbiased estimator of the true global value $V(s)$. This result provides theoretical justification for the critic design in our algorithm: by including the states of $\kappa$-hop neighbors (with $\kappa$ chosen such that $\gamma^\kappa$ is small), we ensure that each UAV's value network $V_u$ captures the vast majority of the state features that influence future rewards. The residual error is bounded and becomes negligible as $\kappa$ grows.

\paragraph{Local Policy Gradient and PPO Updates} A related question is how well policy gradients computed from these local value estimates approximate the true policy gradients one would obtain with a centralized value function. In our decentralized approach, each UAV performs a PPO update using an advantage $\hat{A}_u(s_{N_u},a_u)$ derived from its local critic $V_u(s_{N_u})$. Let $\theta_i$ be the parameter vector of agent $i$'s policy $\pi_i$. The true policy gradient for agent $i$ in a centralized training scenario would be
$g_i = \mathbb{E}_{s\sim d_{\pi},\,a_i\sim \pi_i}\!\Big[ \nabla_{\theta_i} \log \pi_i(a_i \mid s) A_i^{\text{global}}(s, a_i)\Big],$
where $A_i^{\text{global}}(s, a_i)$ is the advantage computed using the {global} value function $V(s)$. In contrast, our decentralized method uses a localized advantage $\tilde{A}_i(s_{N_i^\kappa}, a_i)$ based on $V_i(s_{N_i^\kappa})$. We denote the resulting approximate policy gradient by
$\tilde{g}_i = \mathbb{E}_{s_{N_i^\kappa}\sim d_{\pi},\,a_i\sim \pi_i}\!\Big[ \nabla_{\theta_i} \log \pi_i(a_i \mid s_{N_i^\kappa}) \tilde{A}_i(s_{N_i^\kappa}, a_i)\Big].$

Using Theorem~\ref{thm:value-concentration} and the localized policy assumption $\pi_i(a_i\mid s)\approx \pi_i(a_i\mid s_{N_i^\kappa})$, we can bound the gap between centralized and local policy gradients as follows:

\begin{theorem}[Policy Gradient Approximation Error]\label{thm:policy-gradient}
Under the $\kappa$-hop observability model for each agent and assuming Lipschitz continuity of the policy networks, the norm of the difference between the true centralized policy gradient $g_i$ and the local policy gradient $\tilde{g}_i$ for agent $i$ is bounded by:
\begin{equation}
\big| g_i - \tilde{g}_i\big| \le\frac{2r_{\max}g_{\max}}{(1-\gamma)^2}\gamma^{\kappa}
\end{equation}
where $g_{\max}$ is an upper bound on $|\nabla_{\theta_i} \log \pi_i(a_i \mid s)|$ (the gradient of the policy likelihood, assumed uniformly bounded for all states and actions).
\end{theorem}

By Theorem~\ref{thm:policy-gradient}, locally computed policy updates closely approximate centralized ones when $\kappa$ is sufficiently large (i.e., $\gamma^{\kappa}$ is small).

\end{document}